\documentclass[fleqn,usenatbib]{mnras}

\usepackage{newtxtext,newtxmath}
\usepackage[T1]{fontenc}

\DeclareRobustCommand{\VAN}[3]{#2}
\let\VANthebibliography\thebibliography
\def\thebibliography{\DeclareRobustCommand{\VAN}[3]{##3}\VANthebibliography}

\usepackage{graphicx}	% Including figure files
\usepackage{fix-cm}
\usepackage{amsmath}	% Advanced maths commands
\usepackage{float} 
\usepackage{placeins}
\title[Slow stellar halo rotation]{Slow stellar halo rotation as a signature of disc flips and GES-like mergers}

\author[K. Batrakov et al.]{Kirill Batrakov,$^{1}$\thanks{E-mail: kirill.batrakov@durham.ac.uk}
Alis J. Deason,$^{1,2}$
Francesca Fragkoudi,$^{1}$
Thomas Tomlinson,$^{1}$
Azadeh Fattahi$^{3}$ and 
\newauthor Vasily Belokurov$^{4}$
\\
$^{1}$Institute for Computational Cosmology, Department of Physics, Durham University, South Road, Durham DH1 3LE, UK\\
$^{2}$Centre for Extragalactic Astronomy, Department of Physics, Durham University, South Road, Durham DH1 3LE, UK\\
$^{3}$The Oskar Klein Centre, Department of Physics, Stockholm University, Albanova University Center, 106 91 Stockholm, Sweden\\
$^{4}$Institute of Astronomy, University of Cambridge, Madingley Road, Cambridge CB3 0HA, UK
}

\date{Accepted XXX. Received YYY; in original form ZZZ}

\pubyear{\the\year{}}

\begin{document}
\label{firstpage}
\pagerange{\pageref{firstpage}--\pageref{lastpage}}
\maketitle

\begin{abstract}
The stellar halo of the Milky Way (MW) exhibits a weak net prograde rotational signal  ($v_{\phi} \lesssim 25$ km~s$^{-1}$), yet its origin remains unexplained. To investigate this, we use the Auriga cosmological simulation suite of MW-mass haloes to probe the rotation of stellar haloes over the redshift range $z_{\mathrm{red}}=0-2.5$. The rotational signal is found to persist over this redshift range. We show that satellite progenitors of the stellar halo exhibit a non-uniform distribution of infall directions, with a tendency to align with the host disc, albeit with significant scatter. Thus, we attribute the net rotation to the anisotropic accretion of satellites. Haloes that host \textit{Gaia}-Enceladus-Sausage (GES)-like substructures exhibit consistently slower rotational velocity, likely due to the head-on, more radial trajectory of the dominant progenitor in such systems. Haloes whose stellar discs have reorientated by $\geq 90 \degr$ also exhibit slower rotation, likely because the longer dynamical timescales of haloes prevent them from quickly adjusting to disc reorientations. We also report a correlation between the rotation of stellar and dark matter haloes, suggesting a possible common origin driven by anisotropic accretion. We therefore suggest that the MW experienced a disc flip and has a slowly rotating dark matter halo.
\end{abstract}


%%%%%%%%%%%%%%%%%%%%%%%%%%%%%%%%%%%%%%%%%%%%%%%%%%

%%%%%%%%%%%%%%%%% BODY OF PAPER %%%%%%%%%%%%%%%%%%

\section{Introduction}
\label{sec:intro}
In the $\Lambda$CDM paradigm, dark matter (DM) haloes typically have non-zero net rotation, most often referred to as `spin'. At early stages of halo formation, this spin is driven by tidal torques from the non-uniform environment of the forming haloes \citep[e.g.][]{Peebles69, Doroshkevich70, White84}. At later times, the magnitude and orientation of the DM halo spin are modified by subsequent mergers and accretion events \citep[e.g.][]{Bullock01, Peirani04, DOnghia07}. More specifically, \citet{Vitvitska_DM_AM} and \citet{Hetznecker06} found that massive mergers result in a rapid increase in spin magnitude, although minor mergers also have a significant impact owing to their higher frequency. \citet{Vitvitska_DM_AM} found that the spin magnitude gradually decreases when minor mergers are coming from isotropic directions, while \citet{Hetznecker06} concluded that minor mergers arriving from a preferential accretion direction lead to a gradual increase in spin magnitude. \citet{Bett_spin_flip_I, Bett_spin_flip_II} investigated flips in spin orientation, which were found to be relatively common. Major mergers produce flips more often, although most flips are still attributed to minor mergers due to their higher frequency.

Since DM cannot be observed directly, its merger history has to be reconstructed through some visible tracers -- the most efficient of which is the stellar halo, as it is largely composed of stars from accreted satellite debris \citep[e.g.][]{Cooper10, Naidu20}, which retain the memory of their past trajectories \citep[e.g.][]{Helmi00}. The best `laboratory' for such studies is our own Milky Way (MW), as it provides the most detailed view of a stellar halo, from which we can infer the Galaxy's formation history. However, large samples of halo stars are required to draw robust conclusions.

The launch of the \textit{Gaia} space telescope \citep{Gaia16} has enabled astronomers to study the stellar halo of the MW in unprecedented detail, providing astrometric measurements for 1.8 billion stars. One of the most significant discoveries made with \textit{Gaia} is the \textit{Gaia}-Enceladus-Sausage (GES) \citep{Belokurov_GES, Helmi_GES}, which manifests as a highly radially anisotropic substructure in the metal-rich population of the MW stellar halo. This substructure is believed to originate from an ancient merger ($\approx8-11$ Gyr ago) between the MW and a massive satellite. The properties of the GES merger and its impact on the subsequent evolution of the MW have been the focus of many studies in recent years. For example, GES-like mergers and radially-anisotropic substructures in the present day stellar halo have been found to be relatively common in cosmological simulations of MW-mass haloes: \citet{Fattahi_GES} found that roughly $1/3$ of simulated Auriga galaxies \citep{Auriga17} exhibit GES-like features, and \citet{Dillamore22} found the same $1/3$ fraction in the ARTEMIS simulation suite \citep{ARTEMIS}.  \citet{Dillamore_halo_spin} used the  ARTEMIS suite to further show that systems which host a GES-like substructure have consistently lower DM halo spin than those without a GES-like substructure. However, the impact on the stellar halo spin remains largely unexplored.

The \textit{Gaia} telescope has also made significant contributions to studies of stellar halo spin. Studies dedicated to measuring stellar halo rotation in the MW existed prior to \textit{Gaia}, but were largely limited to using only line-of-sight velocities \citep[e.g.][]{Deason11_haloRot, Das16}, which made inference of the net rotational velocity less reliable. Thanks to \textit{Gaia}, astronomers now have access to large samples of stars with proper motions. \citet{Deason17} combined data from the first data release of \textit{Gaia} \citep[DR1,][]{Gaia16} with observations from the Sloan Digital Sky Survey (SDSS), finding that the net rotational velocity is prograde with respect to the MW disc, with a magnitude of $v_{\phi} \sim 5-25$ km s$^{-1}$ and little variation with the Galactocentric radius. This measurement was compared to stellar haloes in the Auriga simulation suite, which span $0 \lesssim v_{\phi} \lesssim 120$ km s$^{-1}$ when all halo stars are considered, and $0 \lesssim v_{\phi} \lesssim 80$ km s $^{-1}$ when only old stars (age $>10$ Gyr) are included. The observed $v_{\phi}$ of the MW stellar halo therefore falls within the range of simulation predictions, albeit towards the lower end of the $v_{\phi}$ distribution. A more recent study by \citet{Li26} combined precise distance and line-of-sight velocity measurements from the DESI Stellar Surveys \citep[described in detail in the appendix of][]{Dey25} with proper motions from the second data release of \textit{Gaia} \citep[DR2,][]{Gaia18}, finding a net stellar halo rotation of $v_{\phi} \sim 14$ km s$^{-1}$, consistent with \citet{Deason17}.

Recently, the rotation of the inner ($r \lesssim 30$ kpc) stellar halo has been investigated in the context of the influence of bars. This effort has included distribution function (DF)-based modelling \citep[e.g.][]{Li_bar_25, Dillamore_bar_23, Dillamore_bar_24} compared with \textit{Gaia} and APOGEE \citep{APOGEE} data, and the analysis of the integrals-of-motion space in cosmological simulations \citep[e.g.][]{Tomlinson26}. These studies demonstrated that bars can transfer additional spin to the inner regions of stellar halo through resonances. However, the rotation of the outer stellar halo, as well as its net rotation on scales of $r\sim 100$ kpc, has not been studied as extensively. \citet{Deason17} and \citet{Li26} showed that the MW stellar halo exhibits a weak net prograde rotational signal, which lies within the range of values predicted by cosmological simulations. However, how this net rotation is acquired and what determines its magnitude remain open questions, as bars are unlikely to induce resonances at distances as large as $r\sim 100$ kpc. Since the stellar halo encodes the accretion history of its host galaxy, its net rotation is likely linked to the past merger activity. To address these questions, we present a thorough investigation of the stellar halo rotation in the Auriga simulation suite.

This paper is organised as follows. In Section~\ref{sec:auriga} we give a brief overview of the Auriga simulation suite. In Section~\ref{sec:res}, we investigate $v_{\phi}$ as a function of radius at the present day and as a function of redshift to trace its temporal evolution. We also consider how GES-like features and disc reorientations (i.e. disc flips) affect the net rotation of the stellar halo. In Section~\ref{sec:disc} we interpret our results and discuss them in the broader context of satellite anisotropy, GES-like substructures, and disc flips. Finally, in Section~\ref{sec:concl} we summarise the findings of this study.

\section{The Auriga Simulations}
\label{sec:auriga}

Auriga is a suite of cosmological magnetohydrodynamical simulations of MW-mass galaxies \citep{Auriga17, Auriga24}. The initial conditions for the simulations are sampled from the dark matter-only simulation suite EAGLE \citep{Schaye_Eagle} by selecting isolated haloes with virial masses in the range $1-2 \times 10^{12}M_{\odot}$. The evolution of these haloes is then resimulated with stellar and gas particles using a comprehensive galaxy formation model. The simulations were performed using the \textsc{arepo} code \citep{Springel_Arepo, Pakmor_Arepo} in a standard $\Lambda$CDM cosmology, with cosmological parameters $\Omega_{\mathrm{m}}=0.307$, $\Omega_{\mathrm{b}}=0.048$, $\Omega_{\mathrm{\Lambda}}=0.693$, and $H_0=67.77$ km s$^{-1}$ Mpc$^{-1}$ taken from \citet{Planck14}. The typical mass-resolution for baryonic particles is $\approx 5\times10^4 M_{\odot}$, and for DM particles is $\approx3\times 10^5 M_{\odot}$.

The full Auriga suite consists of 30 MW-like galaxies. However, we exclude five galaxies (specifically, Auriga haloes number 1, 11, 20, 25, and 30) as they were identified to be interacting at $z_{\mathrm{red}}=0$ by \citet{Fragkoudi25}. Our sample therefore consists of 25 MW-like galaxies.

For each halo in the sample, we define accreted stellar particles as those formed in haloes other than the main progenitor. Stellar particles formed in the main progenitor of the MW-like halo are labelled as in-situ. Haloes were identified as main or non-main by the \textsc{subfind} algorithm \citep{Springel_Subfind}. These definitions follow the work of \citet{Fattahi_GES}. The stellar halo in this work is then defined as consisting of accreted stellar particles. Such an unambiguous definition is not possible in observations. Therefore, observational studies of the MW stellar halo typically introduce spatial cuts based on the vertical distance of stars from the disc plane to minimise contamination by in-situ stars. For example, \citet{Li26} excluded all stars with vertical distances $|z| < 2$ kpc, while \citet{Deason17} excluded all stars at $|z|<4$ kpc. We discuss the impact of such cuts in Section~\ref{sec:res:z0}.

\section{Results}
\label{sec:res}
In this section, we present the results of our analysis using the Auriga simulation suite. Section~\ref{sec:res:z0} presents measurements of the stellar halo rotation at the current epoch ($z_{\mathrm{red}}=0$); Section~\ref{sec:res:z_evol} presents the evolution of the stellar halo rotation up to redshift $z_{\mathrm{red}}=2.5$; Section~\ref{sec:res:GES_flips} presents the connection between the stellar halo rotation and the presence of GES-like substructures, and the occurrence of disc flips.

\subsection{Present day rotation}
\label{sec:res:z0} 

\begin{figure}
	\includegraphics[width=\columnwidth]{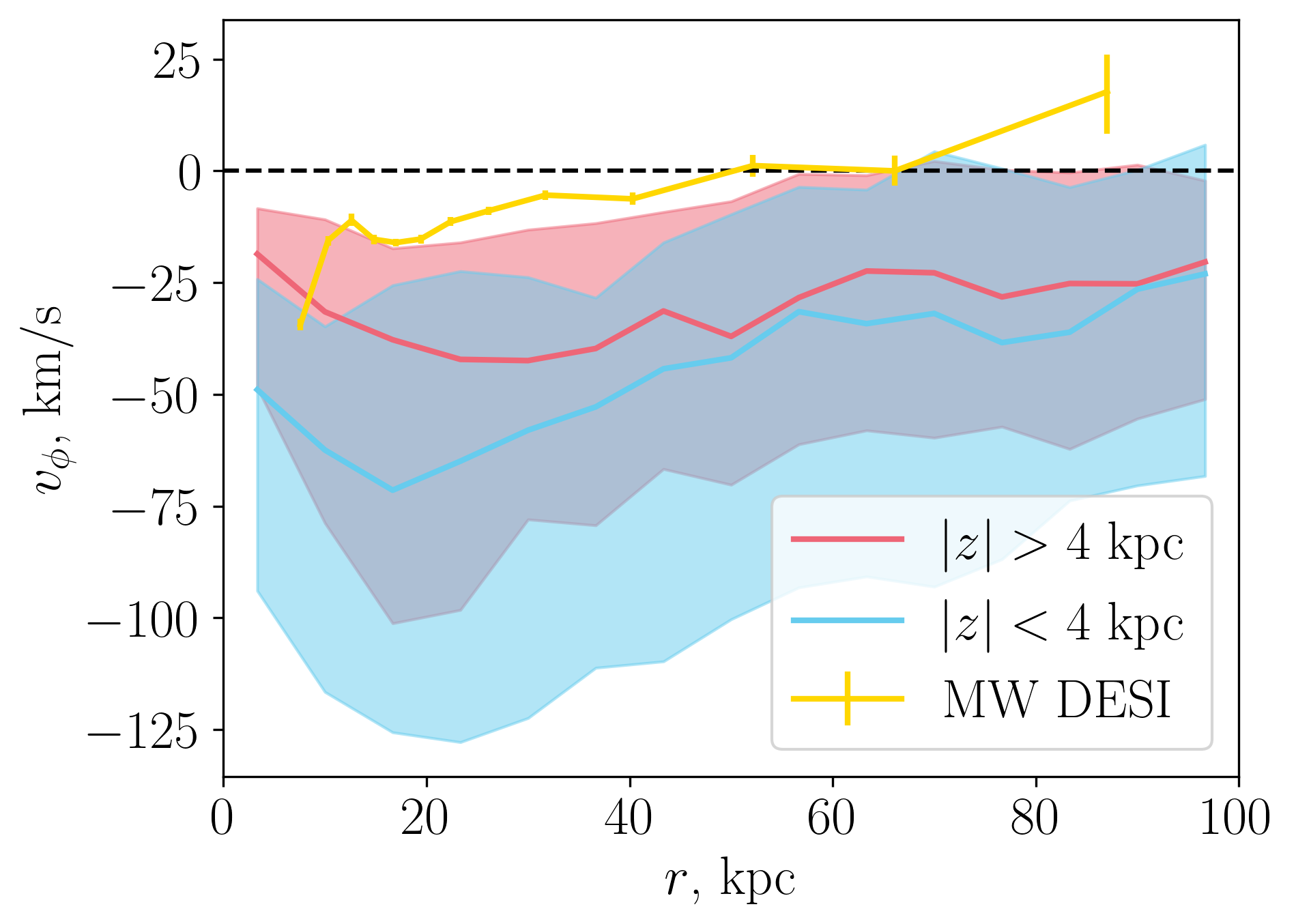}
    \caption[]{Radial profile of the azimuthal velocity component $v_{\phi}$ of stellar haloes at $z_{\mathrm{red}}=0$. The radial range $0-100$ kpc is divided into 15 equally spaced radial bins. Within each bin, the mean $v_{\phi}$ of particles in that bin was measured for each halo. The median, 16th and 84th percentiles (corresponding to $\pm 1\sigma$ around the median) were then computed from the distribution of mean values in each bin. Solid lines show the median, shaded regions show $1\sigma$ intervals.
    Red and green lines show $v_{\phi}$ measured for particles at height $|z|>4$ kpc and $|z|<4$ kpc from the plane of the stellar disc, respectively. $v_{\phi}$ is faster at small radii for particles at height $|z|<4$ kpc due to the `ex-situ discs' reported by \citet{Gomez_exs_discs}. The yellow line shows median $v_{\phi}$ for K-giant stars in the Milky Way stellar halo, as measured by \citet{Li26} using DESI+\textit{Gaia} data; error bars indicate $\pm 1\sigma$ around the median.}
    \label{fig:vphi_radial_Z}
\end{figure}

We use a spherical coordinate system ($r$, $\phi$, $\theta$), where $\phi$ corresponds to the azimuthal direction within the plane of the stellar disc and $\theta$ corresponds to the polar direction perpendicular to the plane of the stellar disc. The azimuthal component of velocity $v_{\phi}$ is therefore a measure of rotation. Note that, following the definition of the $\phi$-direction in our calculations, a negative $v_{\phi}$ represents prograde rotation (with respect to the disc). The general approach for measuring $v_{\phi}$ for the Auriga stellar haloes as a function of radius consists of three steps.
\begin{enumerate}
    \item All accreted stellar particles in a given halo are split into 15 bins equally spaced in $r$ between $0-100$ kpc.
    \item Within each bin, the mean $v_{\phi}$ for each halo is measured, resulting in 25 mean values.
    \item The median of the 25 mean $v_{\phi}$ values is then computed, together with the 16th and 84th percentiles to estimate the $\pm 1\sigma$ spread across all haloes.
\end{enumerate} 
We applied this algorithm to accreted stellar particles at height $|z|<4$ kpc and $|z|>4$ kpc to mimic the spatial cuts usually applied in observational studies. The results of these measurements are shown in green and red, respectively, in Figure~\ref{fig:vphi_radial_Z}.

\begin{figure}
	\includegraphics[width=\columnwidth]{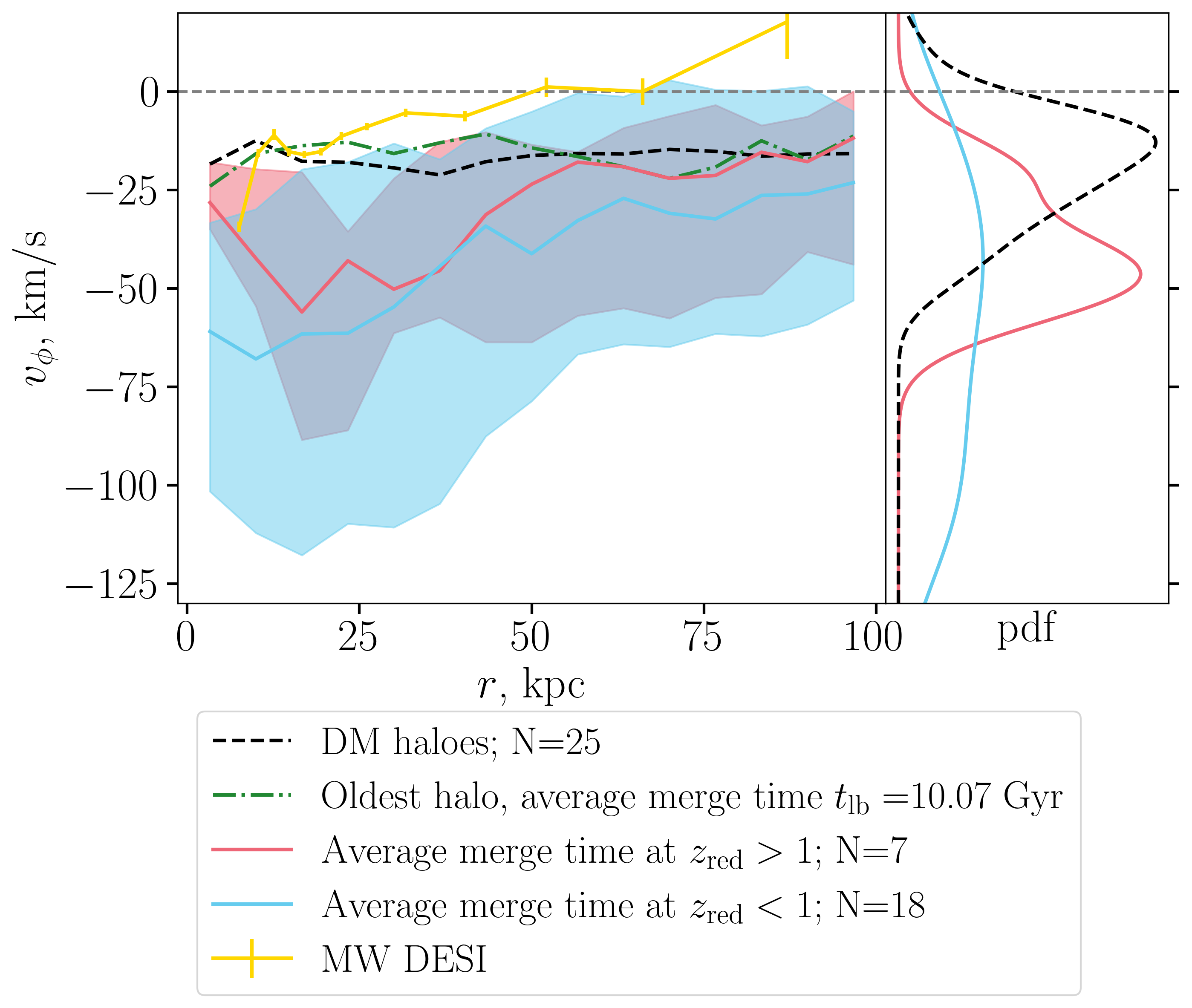}
    \caption[]{Radial profile of the azimuthal velocity component $v_{\phi}$ at $z_{\mathrm{red}}=0$. Solid lines show the medians for two samples of stellar haloes: those with average merger time at $z_{\mathrm{red}}>1$ (earlier-forming, red) and at $z_{\mathrm{red}}<1$ (later-forming, green). The shaded regions show the corresponding $\pm 1\sigma$ intervals. The black dashed line shows the median $v_{\phi}$ for DM haloes. The yellow line shows median $v_{\phi}$ for K-giant stars in the Milky Way stellar halo, as measured by \citet{Li26} using DESI+\textit{Gaia} data; error bars indicate $\pm 1\sigma$ around the median. The panel on the right shows probability distribution functions of the mean $v_{\phi}$ values within $r<100$ kpc for the corresponding categories of Auriga haloes. Later-forming stellar haloes exhibit a broader distribution of mean $v_{\phi}$ compared to earlier-forming stellar haloes and DM haloes.}
    \label{fig:vphi_radial_mergeT}
\end{figure}

\begin{figure}
	\includegraphics[width=\columnwidth]{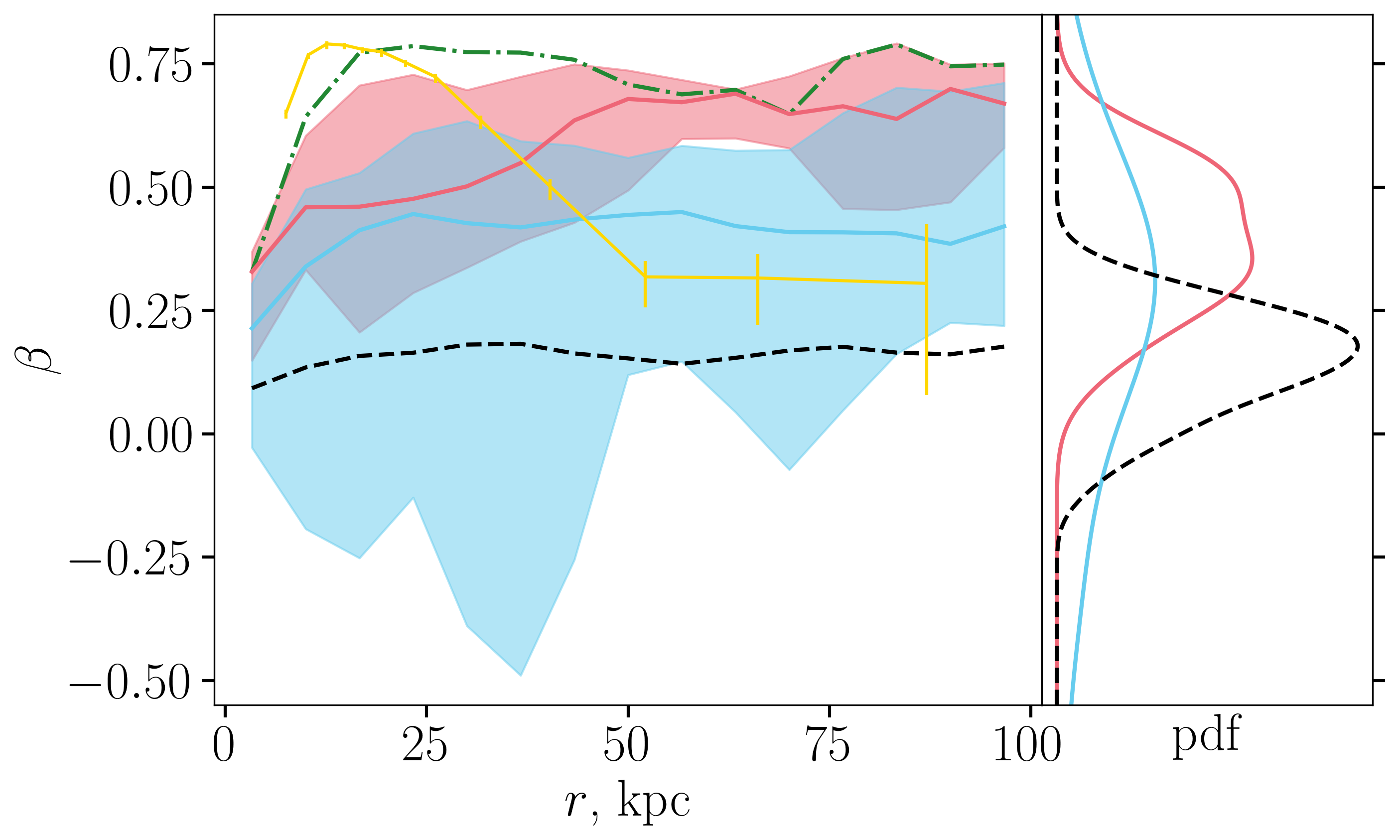}
    \caption[]{Same as Figure~\ref{fig:vphi_radial_mergeT}, but for the radial profile of anisotropy parameter $\beta$. Later-forming stellar haloes again show a broader distribution of mean values, as well as a larger spread in the radial profile and lower median values of $\beta$, corresponding to more circular orbits.}
    \label{fig:beta_radial_mergeT}
\end{figure}

\citet{Gomez_exs_discs} investigated the accreted stellar particles in the Auriga simulation suite that can be kinematically and spatially (i.e., at low heights above the disc plane) attributed to stellar discs. They found that the ratio of the total mass of accreted stellar particles identified as belonging to the disc (termed `ex-situ discs') to the total mass of in-situ stellar particles could be as high as 0.3 for Au20 (excluded from our sample) and 0.15 for Au8 (included in our sample). Approximately one third of all Auriga haloes have a ratio $\geq 0.05$. While ex-situ discs in the other two thirds of the sample were deemed insignificant for studies of stellar discs, they still can have a significant influence on studies of stellar haloes. We demonstrate this in Figure~\ref{fig:vphi_radial_Z}, where the $|z|<4$ kpc selection serves as a rough proxy for `ex-situ discs'. 

We find that stars lying closer to the plane of the disc rotate faster than those located further away. Consequently, samples that include stars at all heights (no $|z|$-cut applied) will be influenced by the faster-rotating stars near the plane, resulting in faster net $v_{\phi}$ than samples restricted to stars above a minimum $|z|$. This means that measurements of $v_{\phi}$ in the Auriga simulation suite that do not impose a spatial $|z|$-cut will result in faster rotational velocities than those found in observational studies that apply such spatial cuts. Figure~\ref{fig:vphi_radial_Z} also shows the radial profile of $v_{\phi}$ in the MW stellar halo as measured by \citet{Li26} using a sample of K-giant stars from DESI+\textit{Gaia} data. Evidently, the MW rotation lies at the lower (closer to zero) boundary of the $1 \sigma$ spread of the Auriga measurements or even above it, depending on the specific radial range considered. Note that, given the relatively small simulation sample size (only 25 haloes) and the rough cut of $|z|>4$ kpc applied in the simulations -- which cannot fully reproduce more sophisticated selection cuts used in observational studies -- it is not surprising that the observational result lies beyond the $1\sigma$ spread of the simulations. Nevertheless, the most important conclusion from Figure~\ref{fig:vphi_radial_Z} is that spatial cuts in $|z|$ affect measurements of $v_{\phi}$. We do not impose such spatial cuts when measuring $v_{\phi}$ at higher redshifts (see Section~\ref{sec:res:z_evol}), as the threshold $|z|>4$ kpc would need to be adjusted at each redshift individually, since the thickness of discs evolves with redshift. Furthermore, the main goal of our study is to investigate which factors lead MW-like galaxies to have relatively slower or faster rotating stellar haloes, while the precise absolute values are of less importance. Thus, no $|z|$-cuts are applied in the subsequent analysis.

\citet{Deason17} found that old (those with age $>10$ Gyr at $z_{\mathrm{red}}=0$) accreted stars in the simulated haloes of the Auriga suite exhibit slower rotation compared to the measurements over all stars, and linked these old halo stars to early mergers and quiescent accretion histories. We therefore want to investigate how the accretion history influences the present-day rotation. To do this, we measure the average merger time for each halo. These are defined as the mean merger time of all accreted particles in a given halo. Since the masses of stellar particles in Auriga are approximately constant, the number of particles associated with a given progenitor is directly proportional to its mass contribution to the stellar halo. Thus, the mean merger time of accreted particles effectively provides a mass-weighted estimate of the average merger time for each halo. All 25 haloes are then divided into two categories: those with an average merger time at $z_{\mathrm{red}}>1$ (earlier-forming) and those with an average merger time at $z_{\mathrm{red}}<1$ (later-forming). The boundary of $z_{\mathrm{red}}=1$ was chosen because the number of haloes increases significantly immediately after this boundary and a higher boundary of $z_{\mathrm{red}}=2$ would result in only one halo in the earlier-forming category.

The measured radial profiles of $v_{\phi}$ for earlier-forming and later-forming haloes are shown in Figure~\ref{fig:vphi_radial_mergeT}. The MW profile measured by \citet{Li26} is shown for reference, together with the oldest stellar halo (Au9) in the sample. The subpanel on the right shows the KDE-smoothed probability distribution function (pdf) curves for mean $v_{\phi}$ for each halo (where we consider all particles within $r<100$ kpc). Evidently, the $v_{\phi}$ curve for earlier-forming haloes (red, $z_{\mathrm{red}}>1$) is slower than for later-forming haloes (green, $z_{\mathrm{red}}<1$). Later-forming haloes also exhibit a significantly broader distribution in $v_{\phi}$ than earlier-forming haloes, with their pdf curves nearly flat over the considered $v_{\phi}$-range. In contrast, earlier-forming haloes display a much narrower and more pronounced peak. This indicates that earlier-forming haloes tend to rotate more slowly, whereas later-forming haloes span a wider range of rotation velocities. 
For comparison, we also show the median $v_{\phi}$ profile of the DM haloes; these exhibit rotation that is even slower than that of earlier-forming stellar haloes. Note that the weak retrograde rotation seen at large $r$ in the MW data is hypothesised by \citet{Li26} to originate from merger events, however specific scenarios are not elaborated.

The kinematic properties of stellar haloes are often analysed by considering their velocity anisotropy, using the $\beta$ parameter defined by \citet{Binney_Tremaine}:
\begin{equation}
    \beta = 1- \frac{\sigma_{\phi}^2 + \sigma_{\theta}^2}{2 \sigma_{r}^2}.
	\label{eq:beta}
\end{equation}
Here, $\sigma_{\phi}$, $\sigma_{\theta}$, and $\sigma_{r}$ are velocity dispersions in spherical coordinates. Note that $\beta=1$ corresponds to a perfectly radial distribution of velocities; $\beta=0$ corresponds to a perfectly isotropic distribution of velocities; $\beta=- \infty$ corresponds to a perfectly circular distribution of velocities. Following the same procedure that we used to measure the radial profile of $v_\phi$, we also measure the radial profile of the anisotropy parameter $\beta$, shown in Figure~\ref{fig:beta_radial_mergeT}. As could be expected from Figure~\ref{fig:vphi_radial_mergeT}, stellar haloes with slower $v_{\phi}$ exhibit higher $\beta$ (i.e., stronger radial anisotropy), while stellar haloes with faster $v_{\phi} $ exhibit lower radial anisotropy. Interestingly, DM haloes, which were previously found to have slower $v_{\phi}$ than any stellar halo category, now exhibit more isotropic profiles than both stellar halo categories. Similarly to the case of $v_{\phi}(r)$, the rapid decrease in $\beta(r)$ seen in the MW data presented by \citet{Li26} is hypothesised to be a result of a complex merger history. It may be due to the majority of stars deposited by the GES merger (which exhibit highly radial orbits and hence high values of $\beta$) being mostly concentrated in the inner regions of the MW stellar halo.
 
\begin{figure*}
    \begin{tabular}{ccc}
\includegraphics[width=0.305\textwidth]{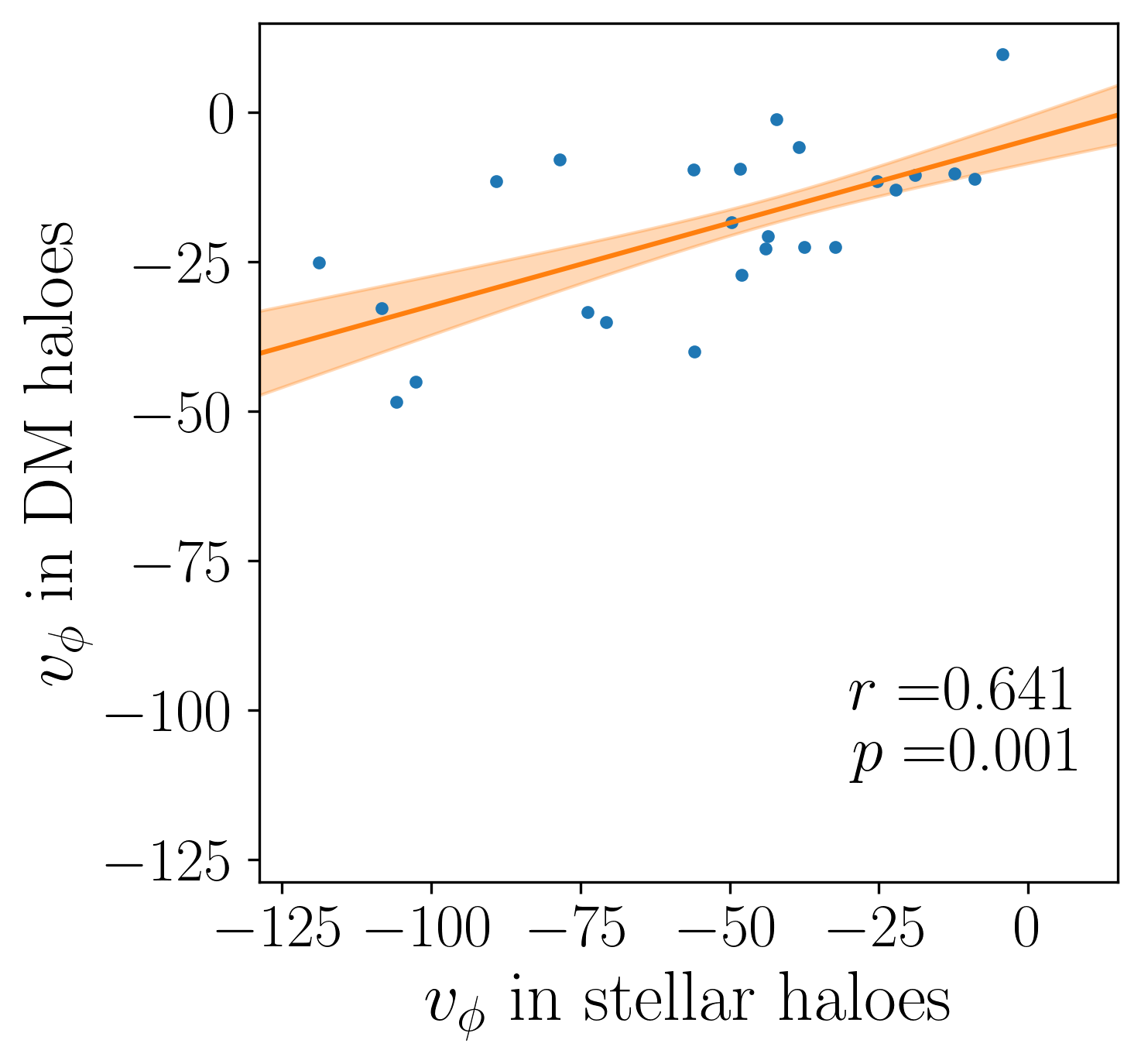} &
\includegraphics[width=0.31021902806\textwidth]{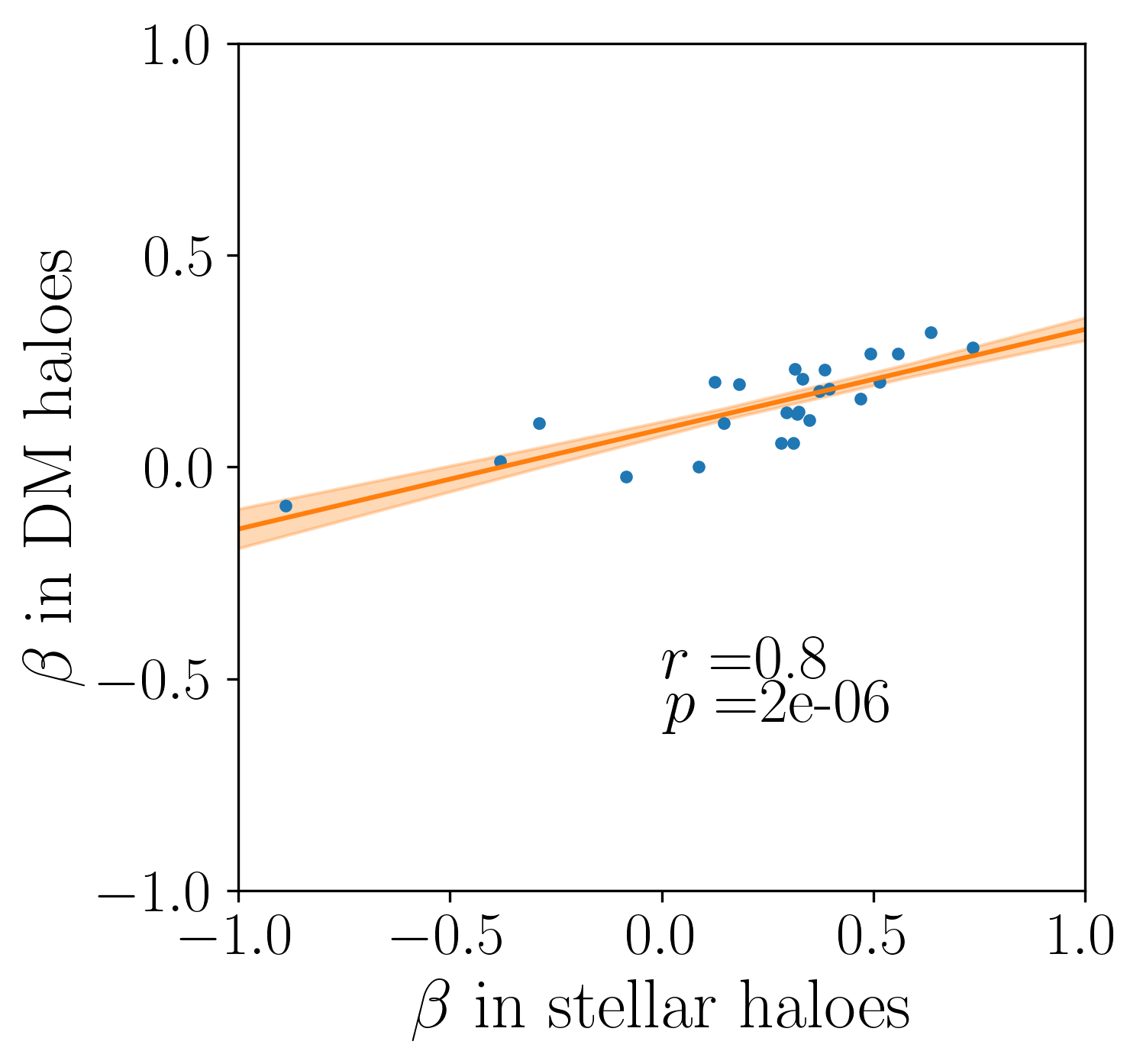} &
\includegraphics[width=0.305\textwidth]{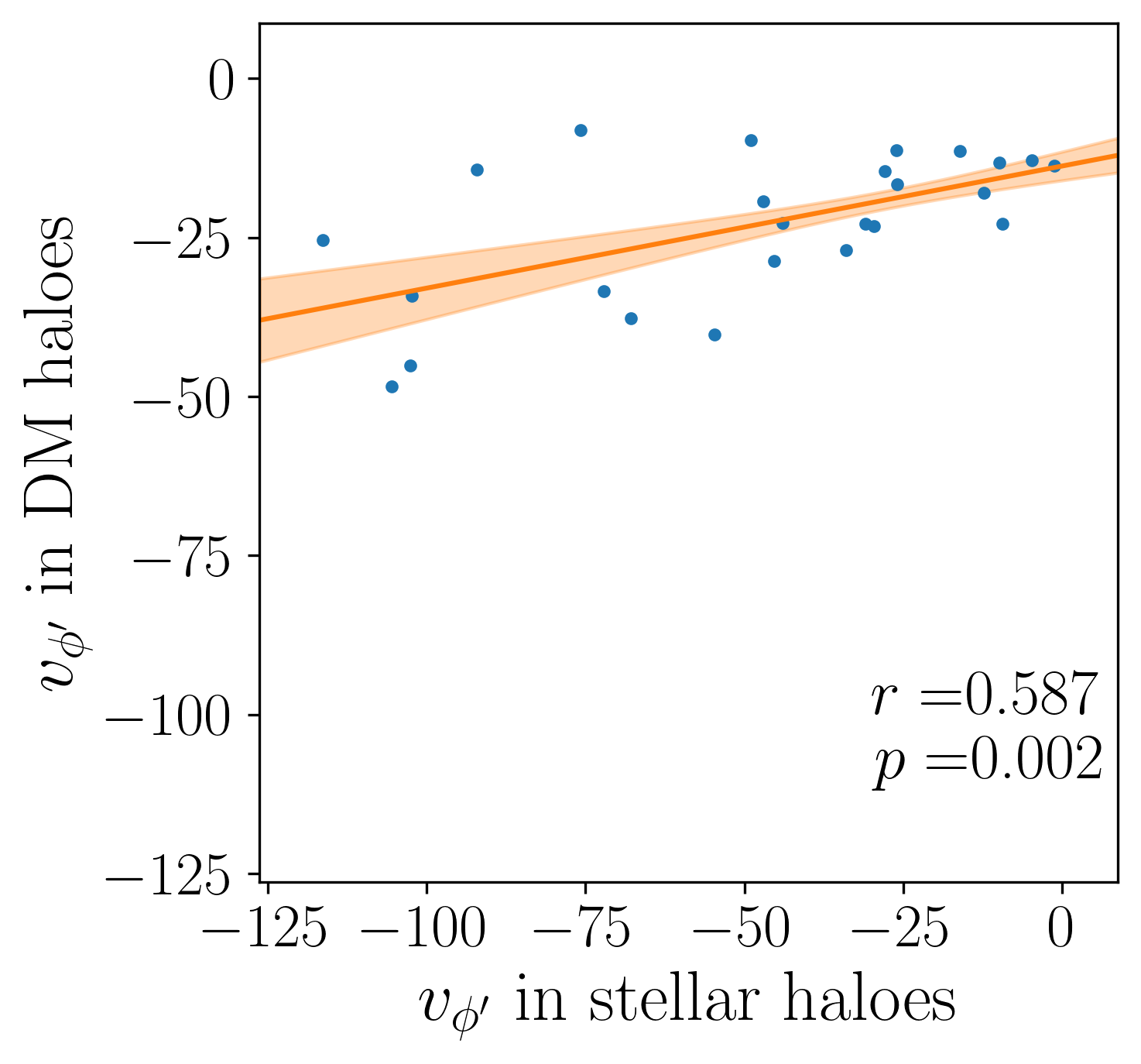} \\
(a) & (b) & (c)
    \end{tabular}
\caption[]{Scatter plots of the mean azimuthal velocity component $v_{\phi}$ with respect to the stellar disc (a), the mean anisotropy parameter $\beta$ (b), and the mean $v_{\phi^{\prime}}$ with respect to the DM reference frame (c), for stellar and dark haloes in Auriga. All $v_{\phi}$, $\beta$, and $v_{\phi^{\prime}}$ are computed for particles within $r<100$ kpc. The straight lines show linear fits. The shaded regions show the uncertainties of fit obtained through jackknife resampling. Pearson correlation coefficients $r$ and corresponding $p$-values are shown below the data-points.}
\label{fig:halo_dm_corr}
 \end{figure*}

The behaviours of DM and stellar haloes differ in Figures~\ref{fig:vphi_radial_mergeT} and \ref{fig:beta_radial_mergeT}. To investigate whether their kinematics are connected, we measure the mean $v_{\phi}$ and mean $\beta$ within $r<100$ kpc for accreted stellar particles and DM particles in each Auriga halo included in our sample. Figure~\ref{fig:halo_dm_corr} shows scatter plots of the mean $v_{\phi}$ in panel (a) and the mean $\beta$ in panel (b). Each point corresponds to an individual Auriga halo, with the measurements for the stellar haloes along the horizontal axis and the measurements for DM haloes along the vertical axis. We perform a linear fit of the form $X_{\mathrm{DM}}=k X_{\mathrm{star}} + b$ for each panel, where $X_{\mathrm{star}}$ and $X_{\mathrm{DM}}$ denote the stellar and DM quantities, respectively. The fits are shown as orange lines in Figure~\ref{fig:halo_dm_corr}. Uncertainties of the fit, shown as shaded regions, are estimated through jackknife resampling. For the rotational velocity $v_{\phi}$, the best-fitting slope is $k=0.278 \pm 0.078$ and the intercept is $b=-4.580 \pm 3.929$; for the anisotropy parameter $\beta$, the slope is $k=0.236 \pm 0.034$ and the intercept is $b=0.089 \pm 0.016$.

We note that the range of values covered by stellar haloes is significantly wider than that of DM haloes, in both rotational velocity and anisotropy parameter. Nevertheless, a robust correlation is established between DM and stellar haloes: the linear fit for $v_{\phi}$ has a $p$-value of 0.001 and a Pearson $r$-coefficient of 0.641, while the linear fit for $\beta$ has a $p$-value of $2\times 10^{-6}$ and a Pearson $r$-coefficient of 0.8. These findings suggest that the rotation of stellar and DM haloes may share a common origin.

To confirm the robustness of this result, we repeat the measurements in a reference frame defined by the DM halo rotation, where the azimuthal plane is perpendicular to the angular momentum vector measured over DM particles within $r<100$ kpc. We label the azimuthal velocity component in this frame as $v_{\phi^{\prime}}$; its scatter plot is shown in panel (c) of Figure~\ref{fig:halo_dm_corr}. The best-fitting slope is $k=0.192 \pm 0.065$ and the intercept is $b=-13.755 \pm 2.259$, with a Pearson $r$-coefficient of $0.587$ and a $p$-value of $0.002$. Although the correlation for $v_{\phi^{\prime}}$ is somewhat weaker than found for $v_{\phi}$, it remains statistically significant, confirming that the result is not specific to the stellar disc reference frame.

\subsection{Redshift evolution of rotation}
\label{sec:res:z_evol}
From Figures~\ref{fig:vphi_radial_mergeT} and \ref{fig:beta_radial_mergeT}, we find that the present day kinematics of stellar haloes are influenced by the average merger times of their progenitors. Thus, we aim to investigate the evolution of rotation with redshift and look for other factors possibly influencing the mean $v_{\phi}$ of stellar haloes.

To measure the evolution of  $v_{\phi}$ with redshift, we adopt a somewhat different approach to that of Section~\ref{sec:res:z0}.
\begin{enumerate}
    \item The mean $v_{\phi}$ for each particle type (accreted stars/in-situ stars/DM) within $r<100$ kpc (decreased to $r<30$ kpc for in-situ stars) is computed for each Auriga halo at a given redshift, $z_{\mathrm{red}}$, resulting in 25 mean values at each $z_{\mathrm{red}}$.
    \item For each particle type, the median of these 25 mean $v_{\phi}$ values is then computed, together with the 16th and 84th percentiles, to estimate the $\pm 1\sigma$ spread across all haloes at a given $z_{\mathrm{red}}$.
\end{enumerate}
We apply this algorithm to all simulation snapshots with redshift up to $z_{\mathrm{red}}=2.5$ (which corresponds to snapshot 57; the present day corresponds to snapshot 127). As discs change their orientation as galaxies evolve, we compare the evolution of rotation both in the present-day azimuthal direction and azimuthal directions corresponding to disc planes at earlier epochs. We use the label $v_{\phi}^{z=0}$ for measurements made with respect to the disc plane at $z_{\mathrm{red}}=0$, and label $v_{\phi}^{z}$ for measurements made with respect to the disc plane at the corresponding (`current') $z_{\mathrm{red}}$. Another modification is the introduction of normalisation by circular velocity at the virial radius and mass:
\begin{equation}
    v^{\mathrm{circ}}_{200} = \sqrt{ \frac{GM_{200c}}{R_{200c}} }.
    \label{eq:vcirc}
\end{equation}
The virial radius $R_{200c}$ and virial mass $M_{200c}$ are extracted at each snapshot using the \textsc{subfind} algorithm \citep{Springel_Subfind}. This normalisation allows us to track $v_{\phi}(z_{\mathrm{red}})$ evolution in more consistent units, since absolute values of $v_{\phi}$ might increase purely due to the mass growth of the host galaxy.

We measure the evolution of $v_{\phi}$ for three types of particles (in-situ stars, accreted stars, and DM). When plotting $v_{\phi}(z_{\mathrm{red}})$ curves, we apply uniform (`boxcar') filtering over a window of 7 neighbouring data-points for measurements at  $z_{\mathrm{red}} \leq 1.5$ to smooth the curves, as the high snapshot density in this region leads to high-frequency variations that obscure the overall trends. At $z_{\mathrm{red}} \geq 1.5$, fewer snapshots are available, so no filtering is applied. The same procedure is adopted for all subsequent $v_{\phi}(z_{\mathrm{red}})$ plots in this Section.

\begin{figure*}
    \begin{tabular}{cc}
    \includegraphics[width=0.45\textwidth]{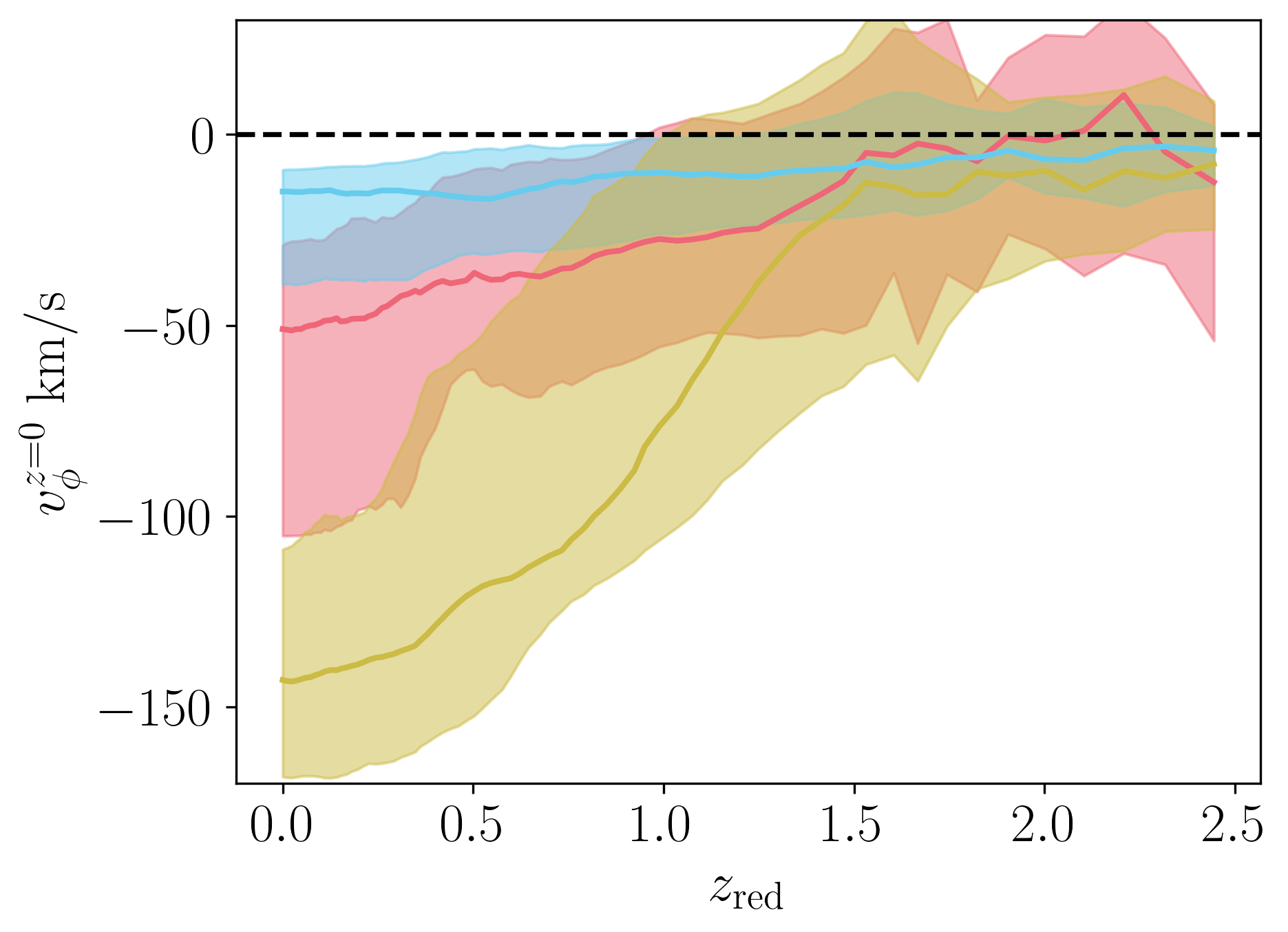} &
    \includegraphics[width=0.45\textwidth]{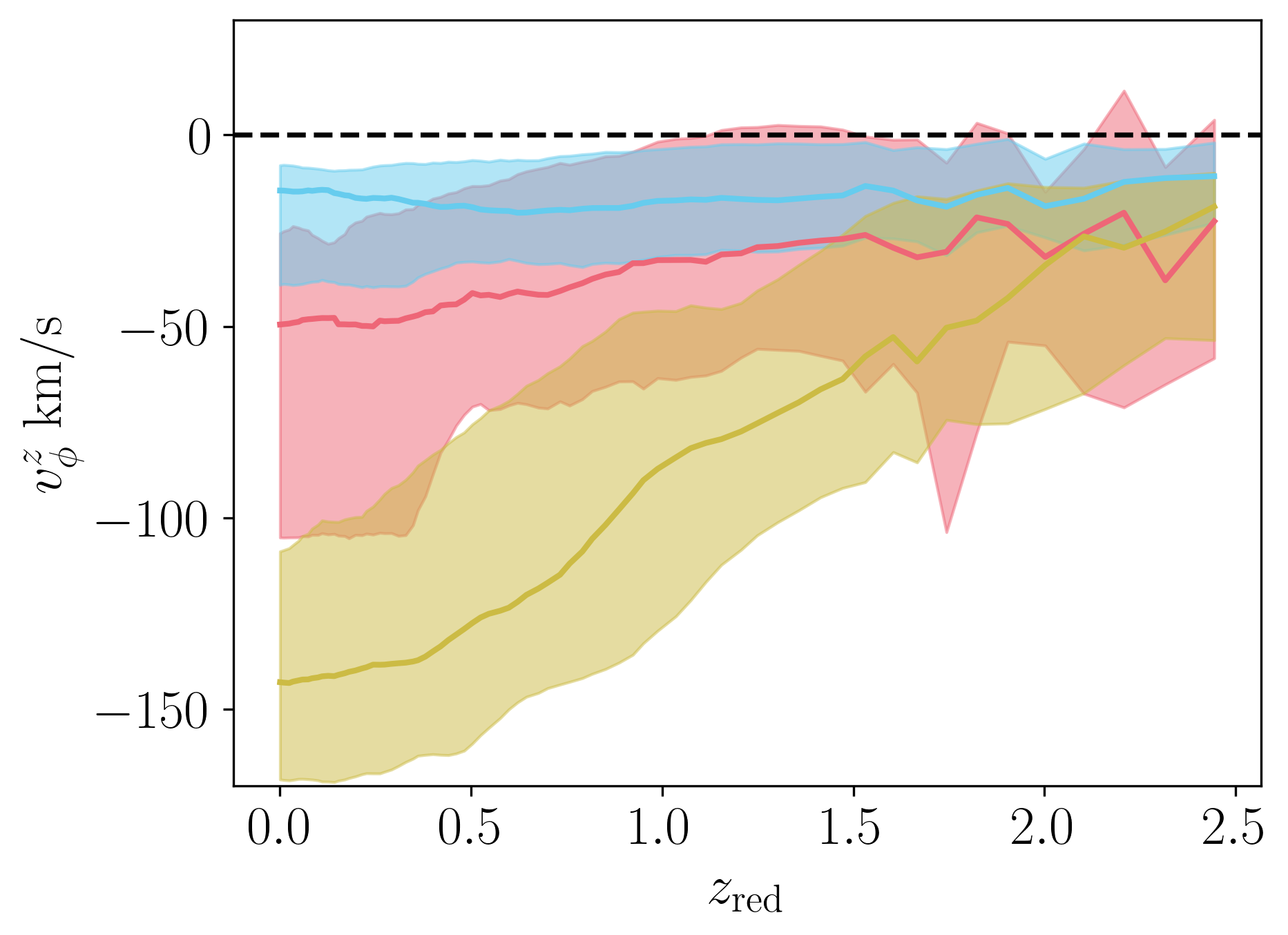} \\
    (a) & (b) \\
    \includegraphics[width=0.45\textwidth]{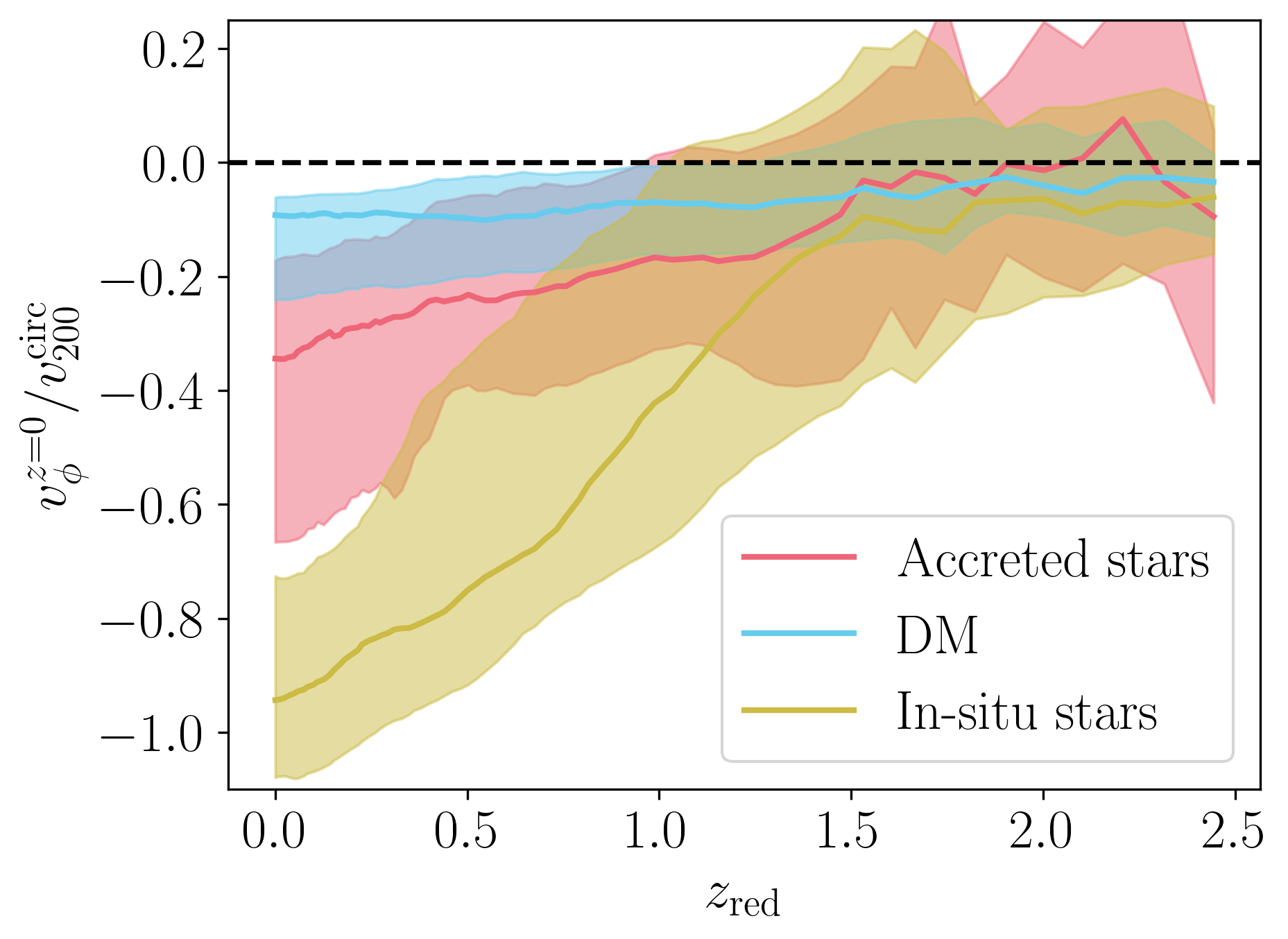} &
    \includegraphics[width=0.45\textwidth]{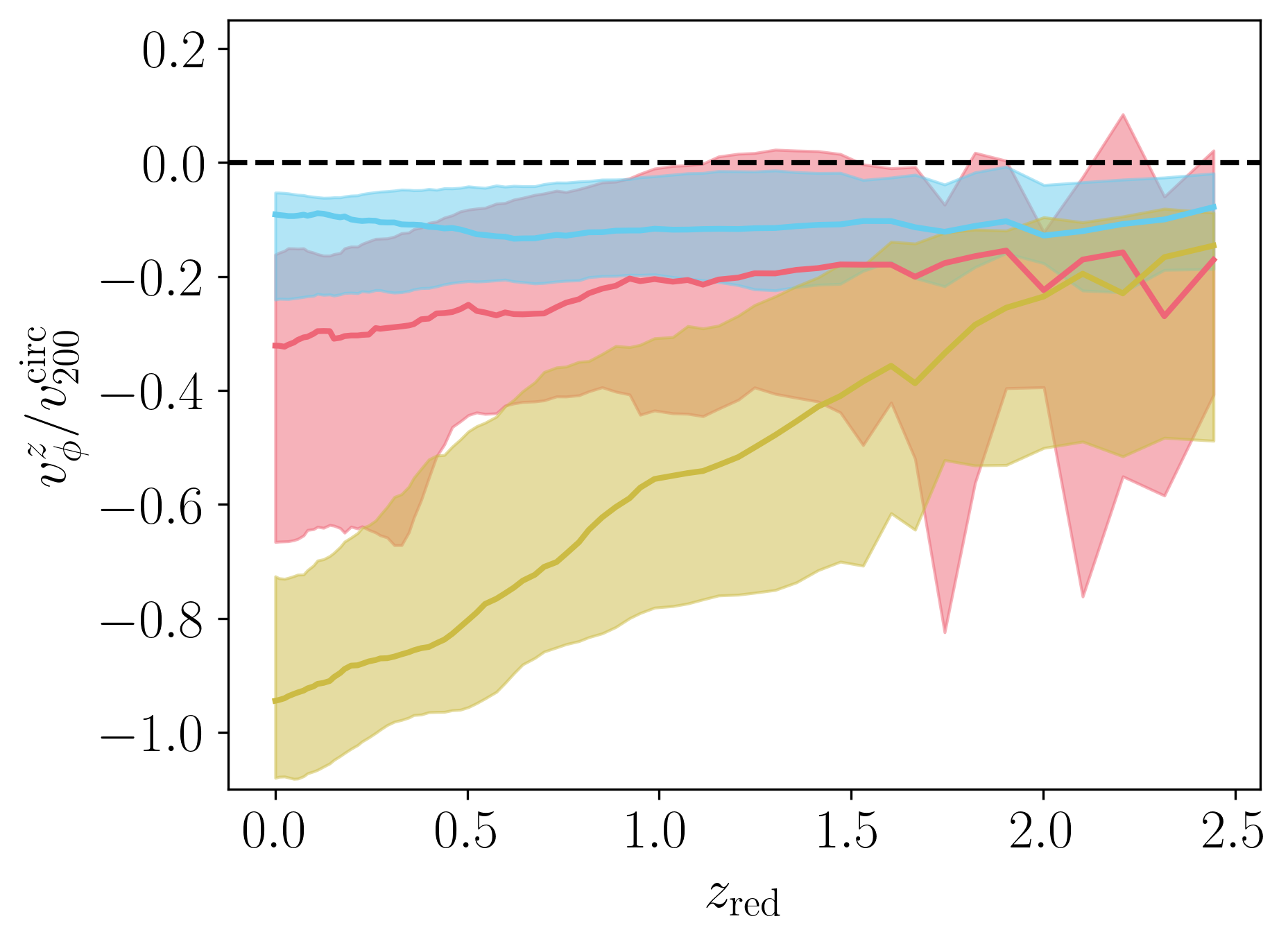} \\
    (c) & (d) 
   \end{tabular}
\caption[]{Evolution of the azimuthal velocity component $v_{\phi}$ for accreted stellar particles (corresponding to stellar haloes; red), in-situ stellar particles (corresponding to stellar discs; yellow), and DM particles (blue) over the redshift range $z_{\mathrm{red}}=0$--$2.5$. As before, solid lines show the medians, and shaded regions show the corresponding $\pm1 \sigma$ intervals. For each halo, a mean $v_{\phi}$ was computed for the respective particle type; the medians were then calculated from the distribution of these mean values. Only particles within $r<100$ kpc and $r<30$ kpc were included for DM/stellar haloes and stellar discs, respectively. Upper panels (a, b) show measurements in absolute values of km s$^{-1}$. Lower panels (c, d) show measurements normalised by the circular velocity $v^{\mathrm{circ}}_{200}$ at virial radius and virial mass. The left panels (a, c) show measurements in the reference frame defined by the plane of the stellar disc of the host galaxy at $z_{\mathrm{red}}=0$ (hence the superscript $z=0$). The right panels (b, d) show measurements in the reference frame defined by the plane of the stellar disc of the host galaxy at current $z_{\mathrm{red}}$ (hence the superscript $z$). Uniform (`boxcar') filter with the window width of 7 was applied for measurements at $z_{\mathrm{red}} \leq 1.5$ to smooth the curves; no filtering was applied at higher $z_{\mathrm{red}}$. Stellar haloes show mild growth in $v_{\phi}$ with decreasing redshift, while stellar discs experience a more rapid increase, nearly achieving $v^{\mathrm{circ}}_{200}$ at $z_{\mathrm{red}}=0$. Note that on average all three particle types remain corotating at all redshifts.}
\label{fig:acc_ins_dm_Zevol}
 \end{figure*}

The four panels of Figure~\ref{fig:acc_ins_dm_Zevol} show the evolution of $v_{\phi}$ under four variations of the measurement procedure: measured relative to the disc plane at $z_{\mathrm{red}}=0$ or at the corresponding $z_{\mathrm{red}}$, and with or without normalisation by $v^{\mathrm{circ}}_{200}$. In-situ stellar particles, which essentially represent stellar discs, show a prominent increase in the magnitude of median $v_{\phi}$ with decreasing redshift, likely reflecting the process of disc formation. DM particles demonstrate very steady median rotation values with redshift. In panels (b, d) DM haloes show a mild slow down at low redshifts. Accreted stellar particles, which correspond to stellar haloes, show a mild increase in rotational velocity at low $z_{\mathrm{red}}$. Importantly, the median values of $v_{\phi}$ for all particle types remain corotating throughout almost all $z_{\mathrm{red}}$ in all panels. 
The $1 \sigma$ envelopes for all particle types remain within $v_{\phi}<0$ (i.e. prograde) when values are computed with respect to the disc plane at the corresponding $z_{\mathrm{red}}$, as shown in panels (b, d) of Figure~\ref{fig:acc_ins_dm_Zevol}. However, the $1 \sigma$ envelopes can extend to $v_{\phi}>0$ (i.e. retrograde) at high redshifts when $v_{\phi}$ is measured with respect to the disc plane at $z_{\mathrm{red}}=0$, as shown in panels (a, c) of Figure~\ref{fig:acc_ins_dm_Zevol}. This is true even for in-situ stellar particles, which suggests that some Auriga haloes experience changes in disc orientation exceeding $90 \degr$ during their lifetime. This will be discussed in more detail in Section~\ref{sec:res:GES_flips}. Finally, we note that the normalisation by the circular velocity at virial mass and radius does not have a significant effect: the upper panels (a, b) and the lower panels (c, d) of Figure~\ref{fig:acc_ins_dm_Zevol} exhibit essentially identical trends.

\begin{figure*}
    %\centering
    \begin{tabular}{cc}
    \includegraphics[width=0.45\textwidth]{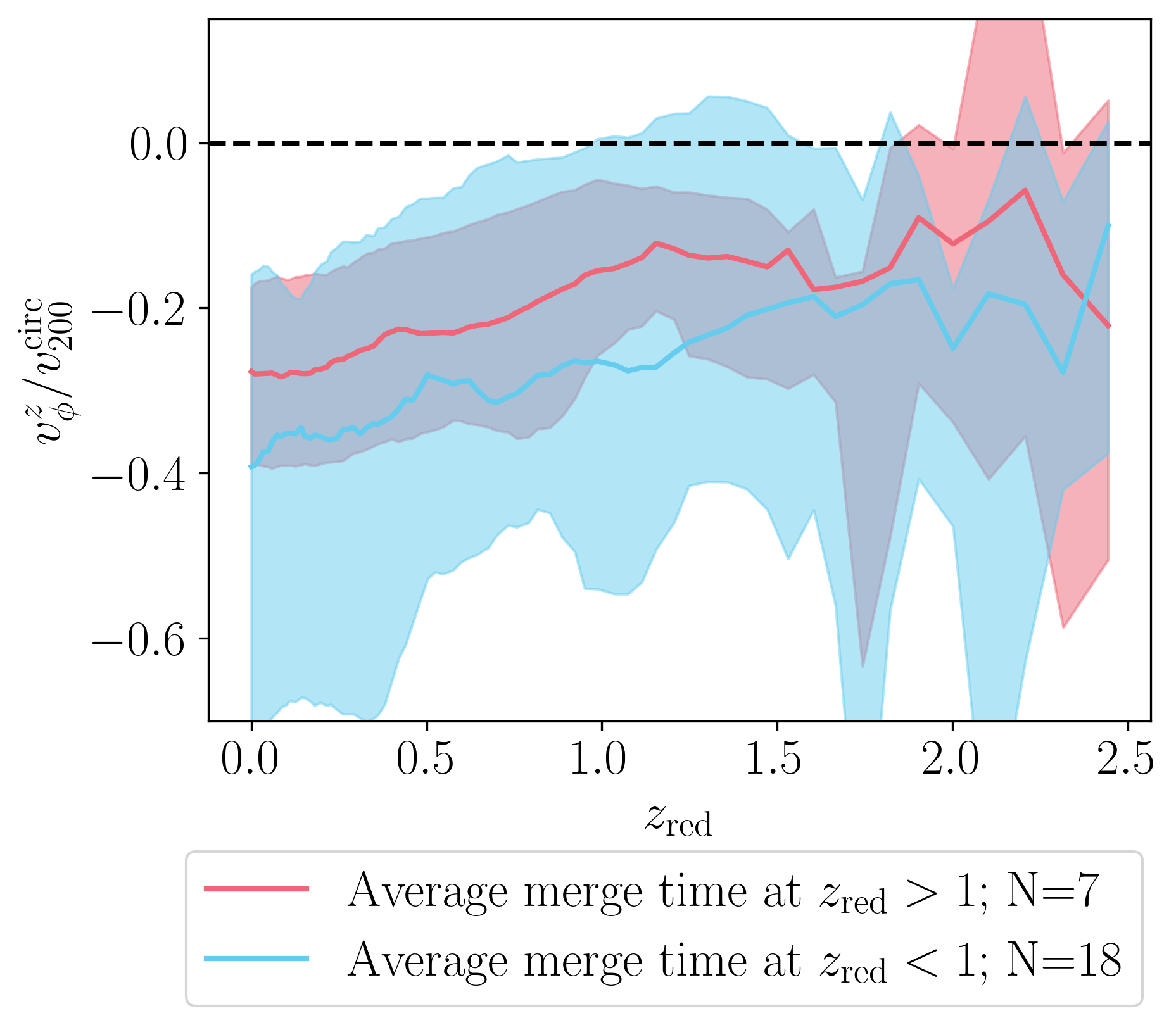} &
    \includegraphics[width=0.45\textwidth]{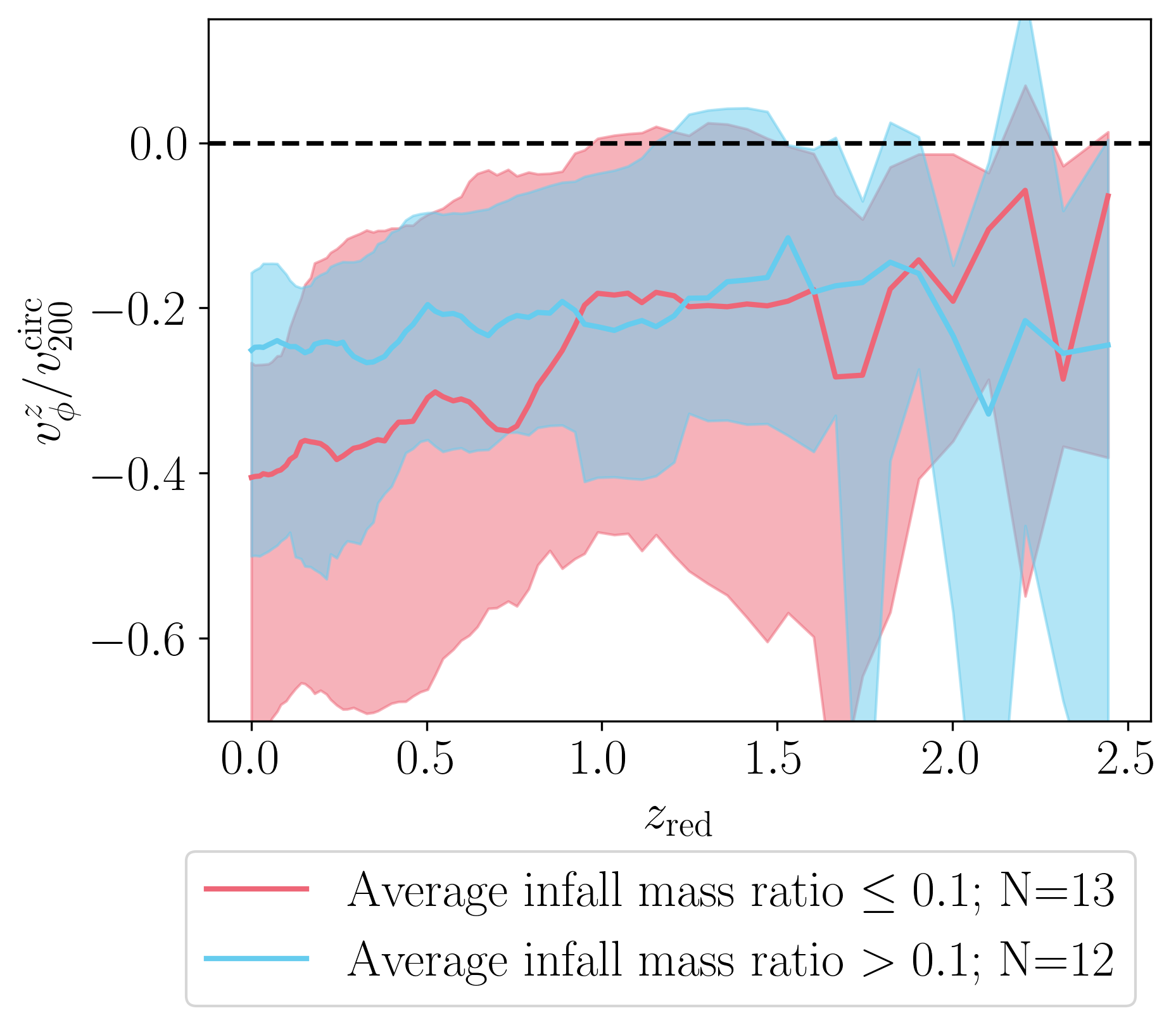} \\
    (a) & (b)
   \end{tabular}
\caption[]{Evolution of the azimuthal velocity component $v_{\phi}$ in stellar haloes split by their average merger time (a) and by their average mass ratio of accreted satellites (b). Solid lines show the medians within the respective categories, and shaded regions indicate the corresponding $\pm 1 \sigma$ intervals. In both panels, $v_{\phi}$ is measured in the reference frame of the stellar disc at current $z_{\mathrm{red}}$ and is normalised by the circular velocity at the virial radius, $v^{\mathrm{circ}}_{200}$.}
\label{fig:acc_mt_Mratio_Zevol}
 \end{figure*}

Following the findings of Section~\ref{sec:res:z0} regarding the importance of the average merger time for present-day rotation, we again split the Auriga haloes into two categories: those that have average merger times at $z_{\mathrm{red}}<1$ (later-forming) and those that have average merger times at $z_{\mathrm{red}}>1$ (earlier-forming).
The redshift evolution of the rotation of stellar haloes of these two categories is shown in panel (a) of Figure~\ref{fig:acc_mt_Mratio_Zevol}. As before, later-forming haloes tend to rotate faster than earlier-forming ones, although this trend is less clear at higher redshifts. More importantly, the broad spread in $v_{\phi}$ observed at $z_{\mathrm{red}}=0$ appears to develop relatively recently, at $z_{\mathrm{red}} \lesssim 0.5$. 

To investigate this further, we identify the four fastest-rotating Auriga stellar haloes at $z_{\mathrm{red}}=0$ with $|v_{\phi}| >100$ km s$^{-1}$ (Au2, Au8, Au16, and Au17) and examine their merger histories. We find that all of them are undergoing some mergers at the present day, with merging satellites having total bound masses of $0.02-0.19$ of the total bound mass of the host galaxy at their corresponding infall snapshots (defined as the first snapshot at which a given satellite crosses the host's virial radius for the first time). This suggests that the significant increase in the $v_{\phi}$ spread at low $z_{\mathrm{red}}$ is being driven by recent mergers.

\cite{Amorisco17} (see also \citealt{Vasiliev_radialization}) reported that the mass ratio of satellites to their hosts is one of the most important parameters defining the orbits of accreted stars. Using idealised N-body simulations, they found that stars originating from more massive satellites exhibit more radial orbits. This potentially implies that mergers with more massive satellites could result in slower rotation. To test this, we identify all merged satellites that contributed at least 100 stellar particles to the present day stellar halo (typically there are $10-20$ such satellites per halo). We measure the ratio of total bound masses of these satellites to those of their hosts at the corresponding infall snapshots, and label this ratio $q$. For each halo, we compute the average infall mass ratio as a weighted average:
\begin{equation}
    \langle q \rangle = \frac{\sum_i q_i \times M_i^{\mathrm{bound}}}{\sum_i M_i^{\mathrm{bound}}},
    \label{eq:M_ratio_avg}
\end{equation}
where the summation is over all satellites of a given halo; $q_i$ is the mass ratio of the $i$-th satellite and $M_i^{\mathrm{bound}}$ is its total bound mass. We split all haloes into two categories: those with average infall mass ratio $\leq 0.1$ and those with average infall mass ratio $> 0.1$. The redshift evolution of rotation for these two halo categories is shown in panel (b) of Figure~\ref{fig:acc_mt_Mratio_Zevol}. Haloes with a smaller infall mass ratio (shown in red) exhibit slightly faster rotation based on the medians, in line with expectations from \citet{Amorisco17}. However, this trend is not consistent across the full redshift range considered, as the two curves intersect multiple times at $z_{\mathrm{red}}>1$. The $1\sigma$ spread for haloes with smaller infall mass ratios is also larger than for haloes with higher mass ratios. However, overall, these differences are less pronounced than those between haloes with different average merger times.
Thus, in light of the weaker trends across mass-ratio categories, we conclude that the average merger time is a more decisive factor governing present-day rotation.

\subsection{GES-like substructures and disc flips}
\label{sec:res:GES_flips}
As discussed in Section~\ref{sec:intro}, many recent studies of the MW have investigated the nature of GES-like substructures in stellar haloes and the influence of GES-like accretion events on its subsequent evolution. This influence has been found to be significant, and we therefore investigate its connection to the rotation of stellar haloes.

\begin{figure*}
    %\centering
    \begin{tabular}{cc}
    \includegraphics[width=0.45\textwidth]{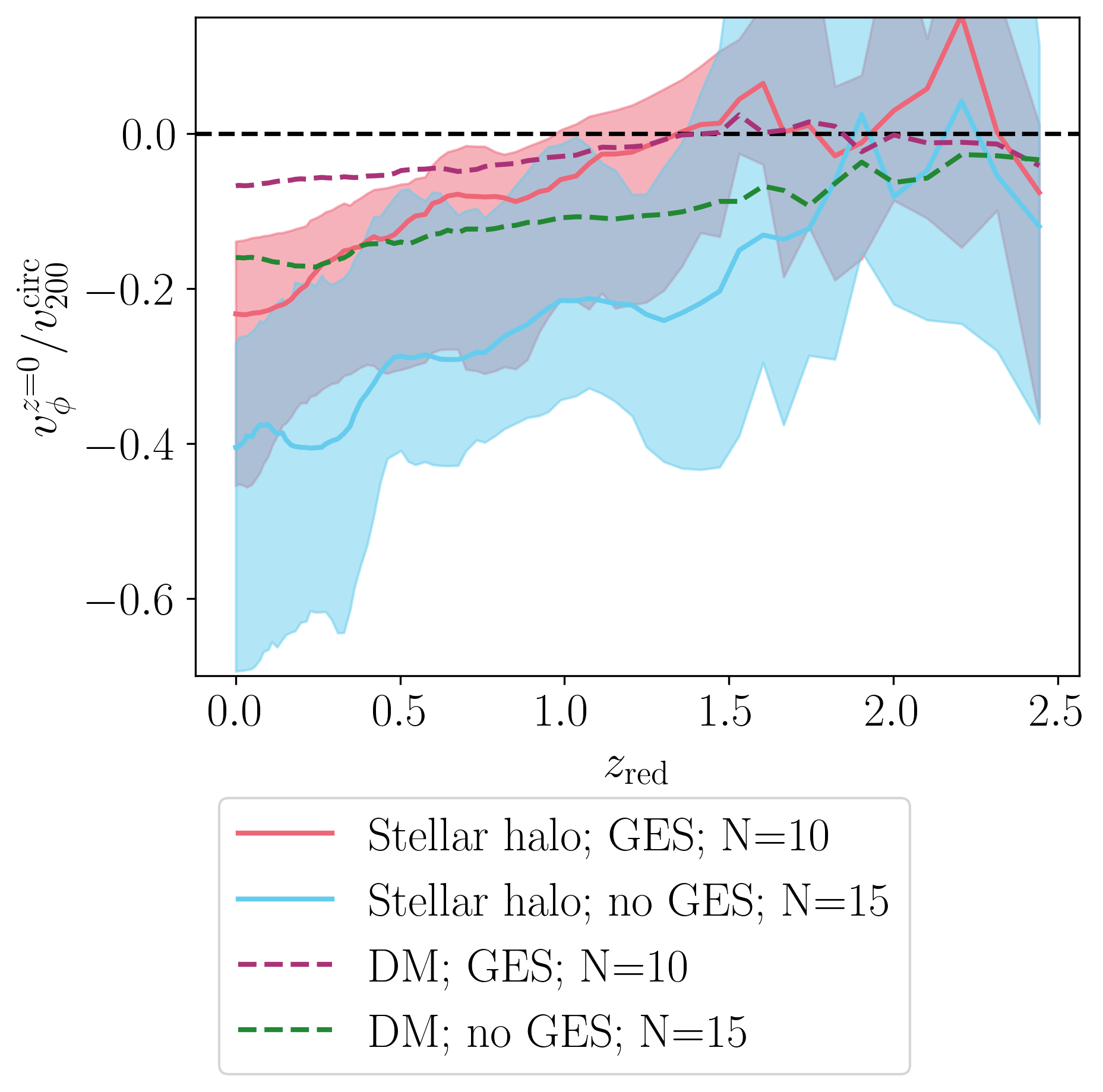} &
    \includegraphics[width=0.45\textwidth]{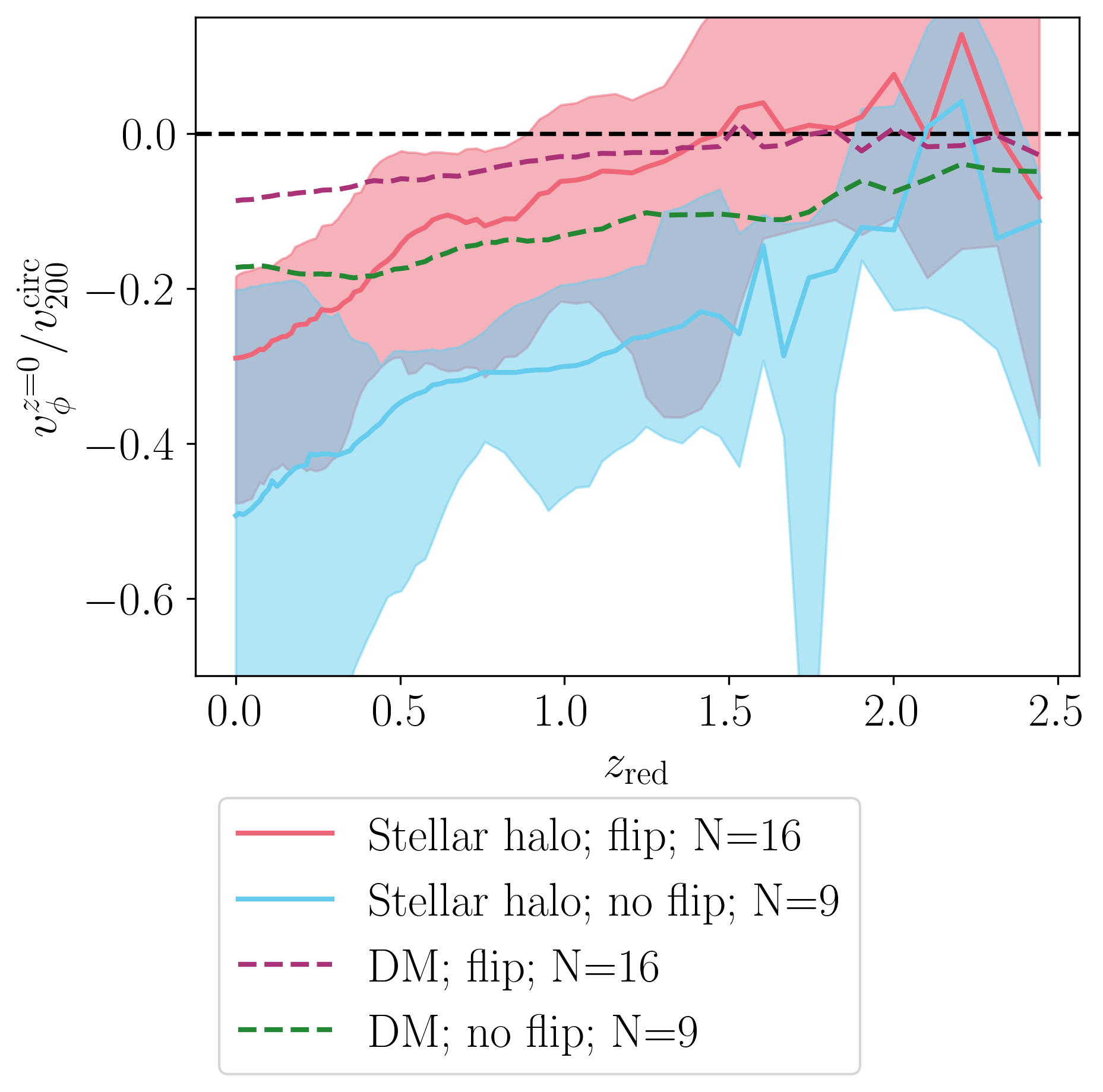} \\
    (a) & (b)
   \end{tabular}
\caption[]{Evolution of the azimuthal velocity component $v_{\phi}$ in stellar haloes split by the presence of a GES-like substructure (a) and by the presence of a disc flip event (b). Solid lines show the medians within the respective categories, and the shaded regions indicate the corresponding $\pm 1 \sigma$ intervals. In both panels, $v_{\phi}$ is measured in the reference frame of the stellar disc at $z_{\mathrm{red}}=0$ and is normalised by the circular velocity at the virial radius, $v^{\mathrm{circ}}_{200}$.}
\label{fig:acc_dm_GES_flip_Zevol}
\end{figure*}

The search for GES-like substructures in Auriga haloes was carried out by \citet{Fattahi_GES}, by identifying stellar haloes exhibiting significantly metal-rich and high velocity anisotropy, $\beta$, populations at the present day. Here, 10 Auriga haloes were identified to host GES-like substructures, all of which are included in our sample. We investigate the $v_{\phi}(z_{\mathrm{red}})$ evolution in haloes with and without a GES-like substructures (for brevity, they are referred to as `GES' and `no GES' in plots, respectively). The procedure is the same as in Section~\ref{sec:res:z_evol} with the only difference being that haloes are split into categories based on the presence of a GES-like substructures. We also trace the $v_{\phi}(z_{\mathrm{red}})$ evolution of DM haloes alongside stellar haloes for those categories. The resulting $v_{\phi}$ vs. redshift profiles are shown in panel (a) of Figure~\ref{fig:acc_dm_GES_flip_Zevol}. Note that in contrast to Figure~\ref{fig:acc_mt_Mratio_Zevol}, the rotational velocity $v_{\phi}$ in Figure~\ref{fig:acc_dm_GES_flip_Zevol} is measured with respect to the disc plane at $z_{\mathrm{red}}=0$. This is because GES-like substructures are identified through the kinematics of stars in present-day haloes, making it more intuitive to trace and analyse the $v_{\phi}(z_{\mathrm{red}})$ evolution in a present-day reference frame. Furthermore, the influence of changes in disc orientation on stellar halo rotation, which we discuss later, is also more clearly seen in a fixed reference frame.

The general trend is that haloes which host a GES-like substructure have consistently slower $v_\phi$ than those which do not host such a substructure. The gap between the medians corresponding to the two categories is wider than in either panel of Figure~\ref{fig:acc_mt_Mratio_Zevol}, indicating that the presence of GES-like features is more important than the average merger time or mass ratio of accreted satellites. The spreads around the medians still overlap substantially, but the narrower distribution (`GES', red) is not fully contained within the wider distribution (`no GES', green), further supporting the significance of GES-like substructures. Interestingly, the median of systems with GES exhibits retrograde rotation (with respect to the present-day disc plane) at high $z_{\mathrm{red}}$, implying that the majority of these haloes likely experienced a disc reorientation of more than $90 \degr$. Furthermore, the rotation of DM haloes in the same categories (GES vs. no GES) is consistent with the trends seen in stellar haloes. The DM haloes exhibit milder magnitudes of $v_{\phi}$, but the DM in systems with GES is consistently slower than the DM in systems without GES. This trend is in agreement with the findings based on the present-day kinematics of haloes in Figure~\ref{fig:halo_dm_corr}.

\begin{figure}
	\includegraphics[width=\columnwidth]{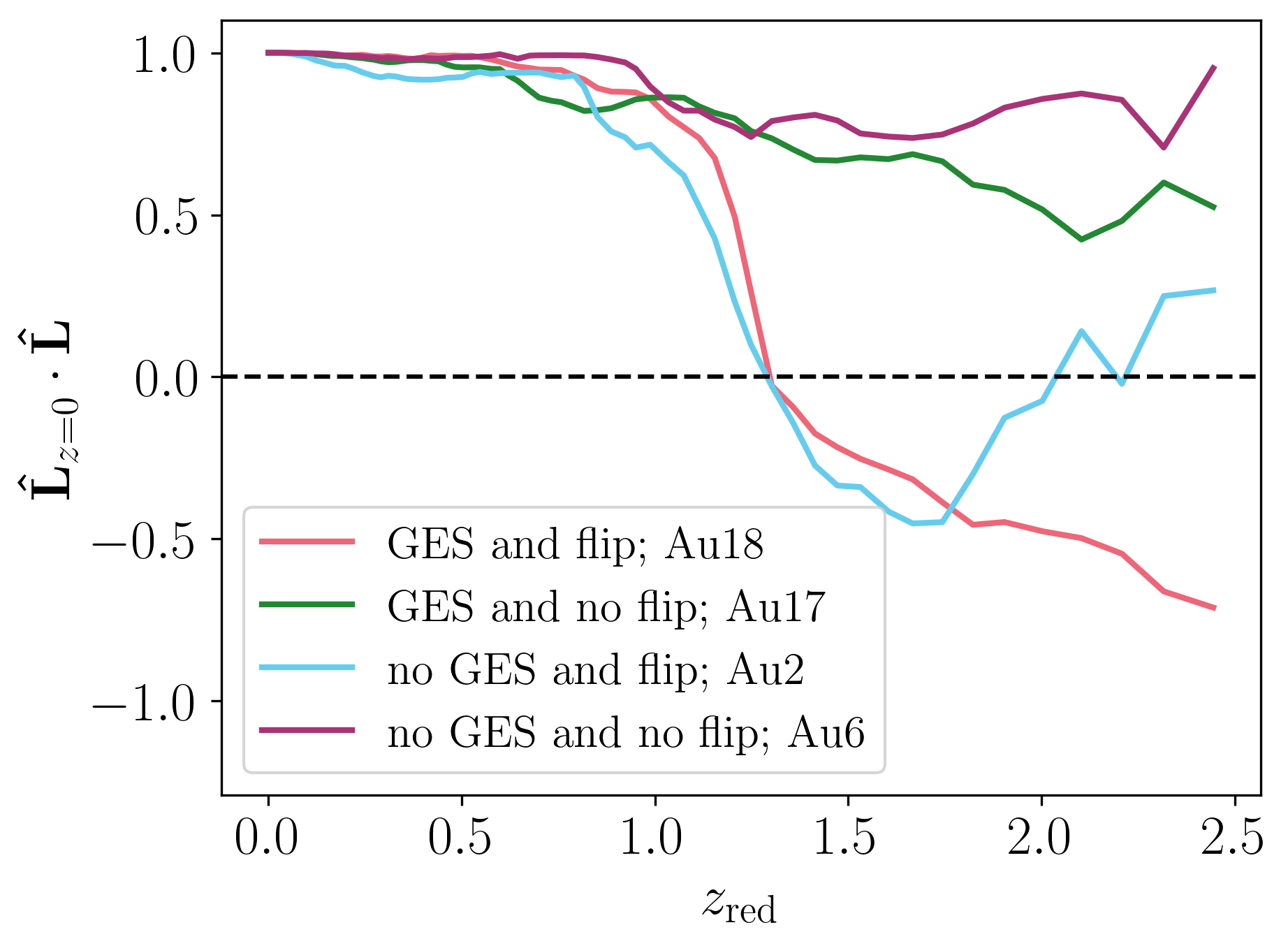}
    \caption[]{Evolution of the orientation of angular momentum (AM) vector of the stellar disc, traced by the dot product of normalised AM vectors at $z_{\mathrm{red}}=0$ ($\mathbf{\hat{L}}_{z=0}$) and at the current $z_{\mathrm{red}}$ ($\mathbf{\hat{L}}$), for four Auriga haloes. These haloes were chosen to represent four categories defined by the presence of a GES-like substructure (as identified by \citealt{Fattahi_GES}) and the presence of a disc flip event (defined here as the change in sign of the dot product shown in this plot).} 
    \label{fig:AM_dot_prod_time_evol}
\end{figure}

As noted previously, some of our figures (e.g. Figure~\ref{fig:acc_ins_dm_Zevol}) suggest that the disc orientation evolves significantly with redshift. \citet{Dillamore22} and \citet{Dodge_disc_tilt} recently reported that stellar discs can undergo substantial tilting as a result of mergers. We therefore investigate further the influence of changes in disc orientation on the evolution of $v_{\phi}(z_{\mathrm{red}})$. At each snapshot across the redshift range considered, we select all in-situ stellar particles within $r<30$ kpc. This selection serves as a proxy for the stellar disc, as disc particles dominate at small radii. We measure the normalised angular momentum (AM) vector of this selection of stellar particles in each halo, labelling the normalised AM vector at $z_{\mathrm{red}}=0$ as $\mathbf{\hat{L}}_{z=0}$ and that at the corresponding (`current') $z_{\mathrm{red}}$ as $\mathbf{\hat{L}}$. We then compute their dot product $\mathbf{\hat{L}}_{z=0} \cdot \mathbf{\hat{L}}$ to track the evolution of the disc orientation. A negative dot product at a given redshift indicates that the disc was misaligned from its $z_{\mathrm{red}}=0$ orientation by more than $90 \degr$ at that time. We label such an event as a `disc flip'. Figure~\ref{fig:AM_dot_prod_time_evol} shows the evolution of the normalised AM vector dot products up to $z_{\mathrm{red}}=2.5$ for four representative Auriga haloes. To verify that the identified changes in AM orientation reflect a genuine reorientation of the disc rather than the formation of a younger counter-rotating disc (see \citealt{Corsini_counter_rot} for a review of this phenomenon), we repeat the measurements of the AM vector orientation considering only old stars (those with age $\geq 10$ Gyr at $z_{\mathrm{red}}=0$).
We confirm that all Auriga haloes identified as having experienced a disc-flip event also exhibit such an event in the old stellar population. This indicates that the formation of younger counter-rotating discs does not play a significant role in driving these events.

We define two new halo categories: those that experienced a disc flip and those that did not. Their respective $v_{\phi}(z_{\mathrm{red}})$ evolution is shown in panel (b) of Figure~\ref{fig:acc_dm_GES_flip_Zevol}. In general, haloes that experienced a disc flip event are rotating slower than those without a disc flip. Notably, for $0.5 \lesssim z_{\mathrm{red}} \lesssim 2.0$, the spreads around the two medians have minimal overlap, further suggesting that the two categories are significantly distinct in their rotational velocities. The rotation of DM haloes is again consistent with their stellar counterparts within the respective categories. Based on our findings, we therefore conclude that the occurrence of a disc flip and the presence of a GES-like substructure are the two most important factors in shaping the present day rotation of stellar haloes.

\subsubsection{Orbital properties of accreted satellites}
\begin{figure*}
    %\centering
    \begin{tabular}{cc}
    \includegraphics[width=0.45\textwidth]{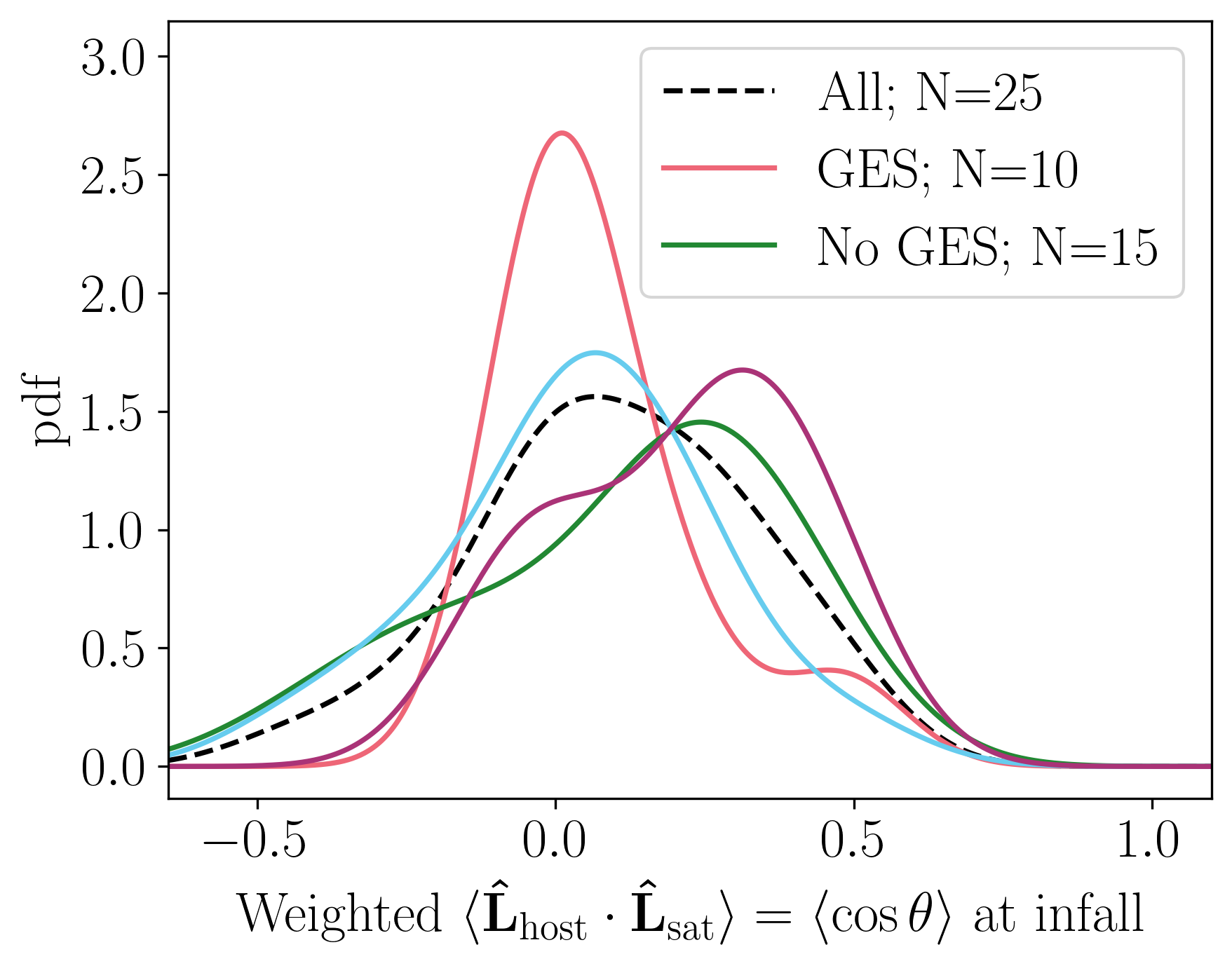} &
    \includegraphics[width=0.44\textwidth]{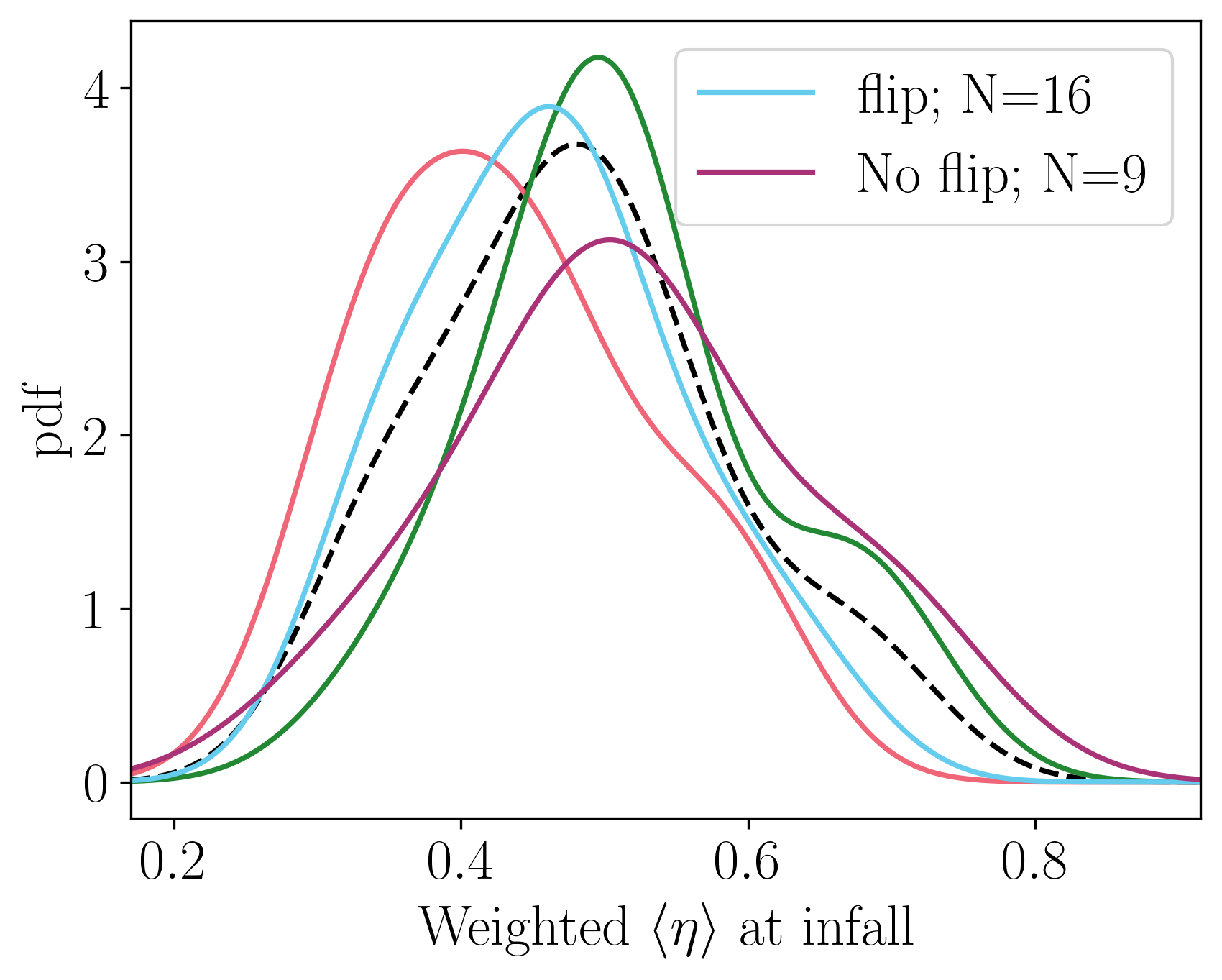} \\
    (a) & (b) 
   \end{tabular}
\caption[]{Smoothed distribution plots for weighted averages of halo progenitors orbital parameters (misalignment angle $\cos \theta$ and circularity $\eta$) for different categories of Auriga haloes. Mass ratios of the satellite to the host at infall were used as weights when averaging. Note that haloes with GES are clearly distinct in both panels.}
\label{fig:eta_theta_kde}
 \end{figure*}

Our next goal is to investigate the connection between the orbital properties of accreted satellites and the presence of GES-like substructures and disc flip occurrences. We use two quantities to characterise orbits: the cosine of misalignment angle $\cos \theta$ and the circularity $\eta$. The former is measured similarly to the disc orientation. We measure the normalised AM vector of the host stellar disc $\mathbf{\hat{L}}_{\mathrm{host}}$ and the normalised AM vector of the satellite's orbit $\mathbf{\hat{L}}_{\mathrm{sat}}$ in the reference frame of the host; both AM vectors are measured at the infall snapshot of the satellite. Then, we compute their dot product, which is the cosine of misalignment angle:
\begin{equation}
    \cos \theta =  \mathbf{\hat{L}}_{\mathrm{host}} \cdot \mathbf{\hat{L}}_{\mathrm{sat}}.
    \label{eq:misalignment}
\end{equation}
The circularity is defined as the ratio of the satellite's tangential orbital velocity component -- measured with respect to the reference plane of the host stellar disc -- to the total orbital velocity modulus:
\begin{equation}
    \eta = \frac{v_{\mathrm{tan}}}{v_{\mathrm{tot}}}.
    \label{eq:circularity}
\end{equation}
These two quantities are measured for all satellites that contributed at least 100 stellar particles to the present-day halo. Measurements were made at infall snapshots.  
We compute mass-weighted averages of $\cos \theta$ and $\eta$ for each halo from their respective satellites, using the ratios of the satellite's total bound mass to the host's total bound mass at the infall snapshot as weights:
\begin{equation}
    \langle \cos \theta \rangle = \frac{\sum_i q_i \times \cos \theta_i}{\sum_i q_i}; \; \;
    \langle \eta \rangle = \frac{\sum_i q_i \times \eta_i}{\sum_i q_i},
    \label{eq:orb_par_avg}
\end{equation}
where the summation is over all satellites of a given halo; $q_i$ is the mass ratio of the $i$-th satellite and $\eta_i$, $\cos \theta_i$ are its orbital parameters. The KDE-smoothed distributions of $\langle \cos \theta \rangle$, and $\langle \eta \rangle$ for five halo categories (all, with GES, without GES, with disc flip, and without disc flip) are shown in Figure~\ref{fig:eta_theta_kde}.

In both panels, it is clear that all categories exhibit a non-uniform distribution of orbital parameters. This implies that satellites, on average, are arriving from some preferred direction, rather than accreting isotropically. Furthermore, the haloes that host a GES-like substructure at the present day are the most distinct from the others. Haloes in this category have their satellites on nearly perpendicular ($\langle \cos \theta \rangle \approx 0$ which corresponds to $\theta \approx 90 \degr$) and significantly radial orbits ($\langle \eta \rangle \lesssim 0.5$) at infall. While the distributions for other halo categories are not identical, their similarity to one another makes it more difficult to draw robust conclusions.

\begin{figure}
\includegraphics[width=0.48\textwidth]{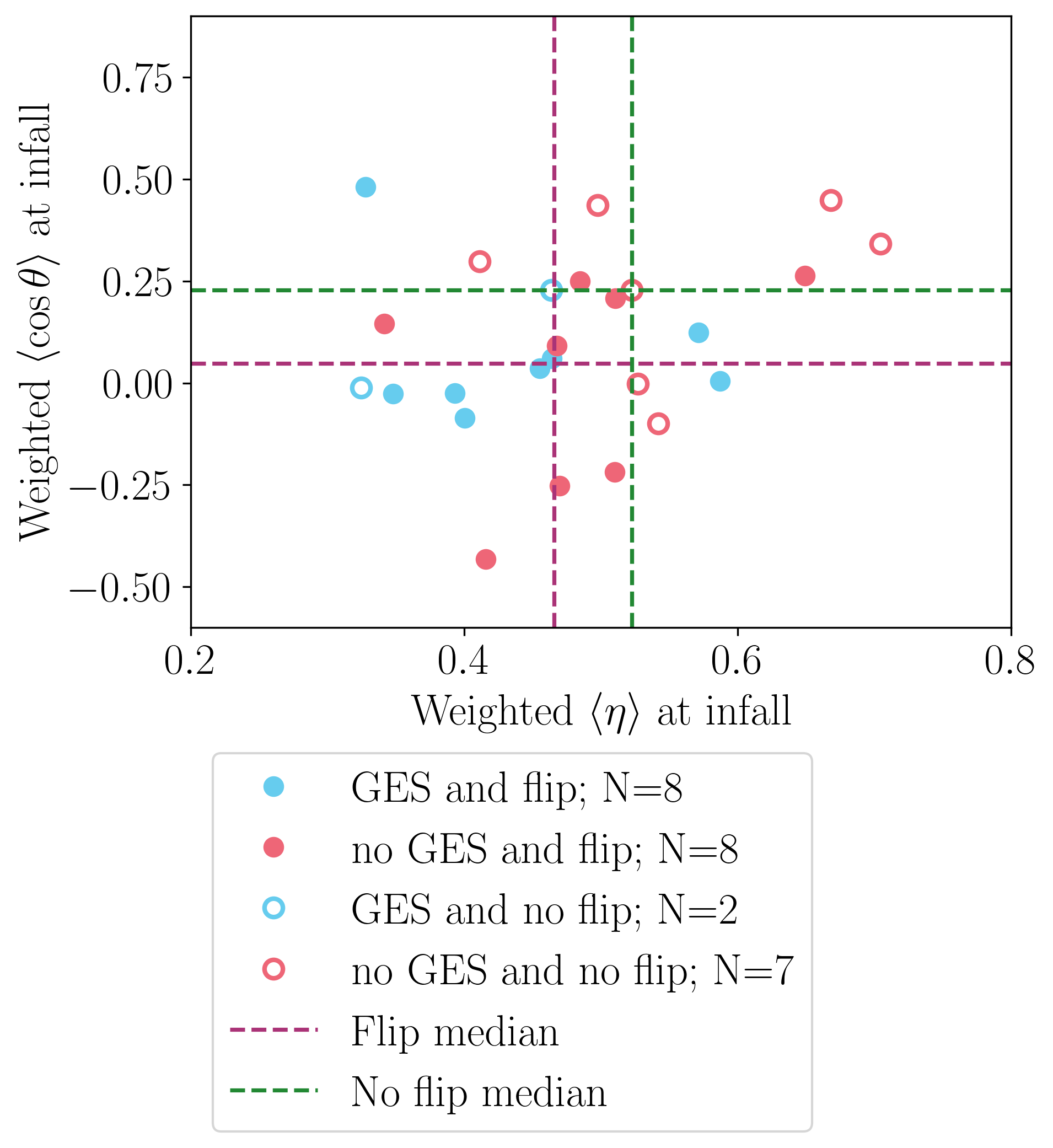}

\caption[]{Scatter plot of weighted averages of circularities $\langle \eta \rangle$ against misalignment angles $\langle \cos \theta \rangle$, for different halo categories split by the presence of a GES-like substructure and a disc flip. Dashed lines show the median values of $\langle \eta \rangle$ and $\langle \cos \theta \rangle$ for haloes with and without disc flips.}
\label{fig:eta_angle}
 \end{figure}

To examine the distribution of haloes in both $\theta$ and $\eta$ simultaneously, we turn to the scatter plot of Figure~\ref{fig:eta_angle}. Filled points correspond to haloes that experienced a disc flip, whereas empty points correspond to haloes without one. Dashed lines show the median values of each parameter: orange for haloes with a disc flip, and purple for haloes without a disc flip. Finally, green points correspond to haloes with GES, whereas red points correspond to haloes without GES. Interestingly, there are only two haloes in the sample that host a GES-like substructure, but did not experience a disc flip. However, many more haloes ($N = 8$) experienced a disc flip without a GES-like substructure.
The general trend is that haloes with disc flips tend to have satellites on orbits with lower $\cos \theta$ (corresponding to orbits that are less aligned with the host stellar disc) at infall, a difference consistent with the result of a Kolmogorov-Smirnov (KS) test ($D=0.49$, $p=0.10$). A similar comparison of the circularity $\eta$ at infall shows the preference of haloes with disc flips for lower $\eta$, albeit a KS test ($D=0.31$, $p=0.34$) does not support the statistical significance of the difference between the medians. We treat these as suggestive trends, as the modest sample size makes it difficult to draw robust conclusions.

 \begin{figure*}
    %\centering
    \begin{tabular}{cc}
\includegraphics[width=0.4\textwidth]{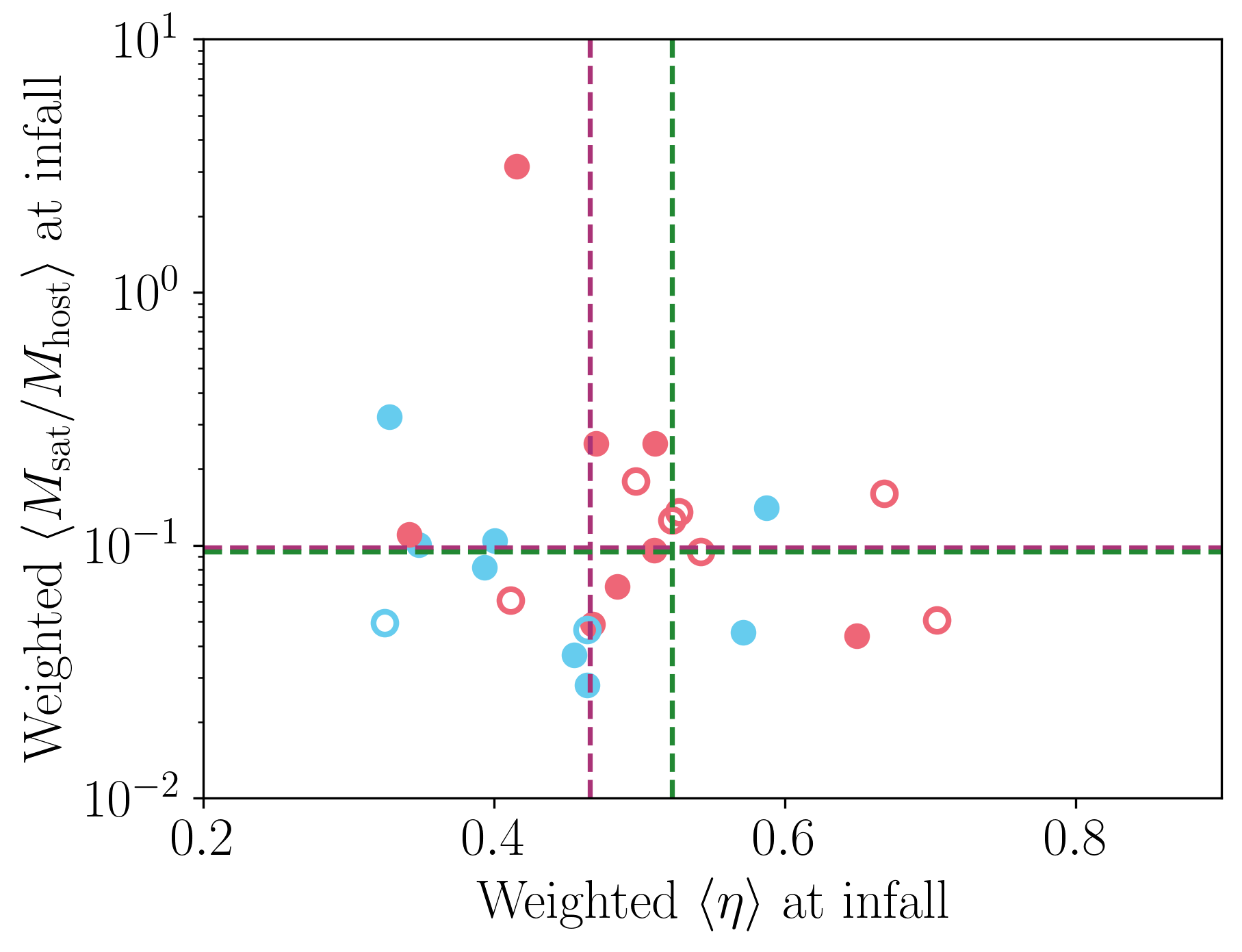}
&
\includegraphics[width=0.422\textwidth]{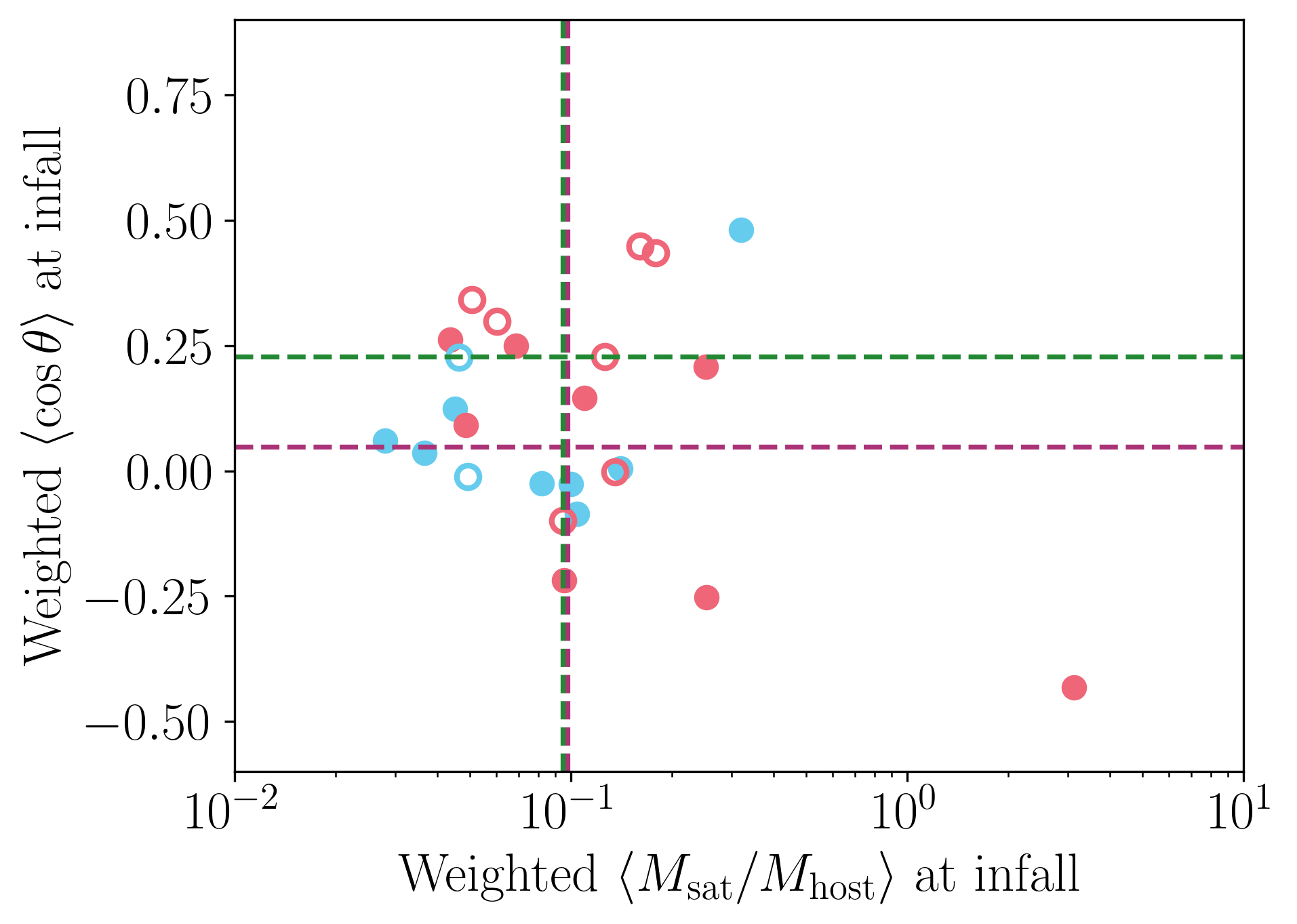} \\
    (a) & (b) \\ 
\includegraphics[width=0.383\textwidth]{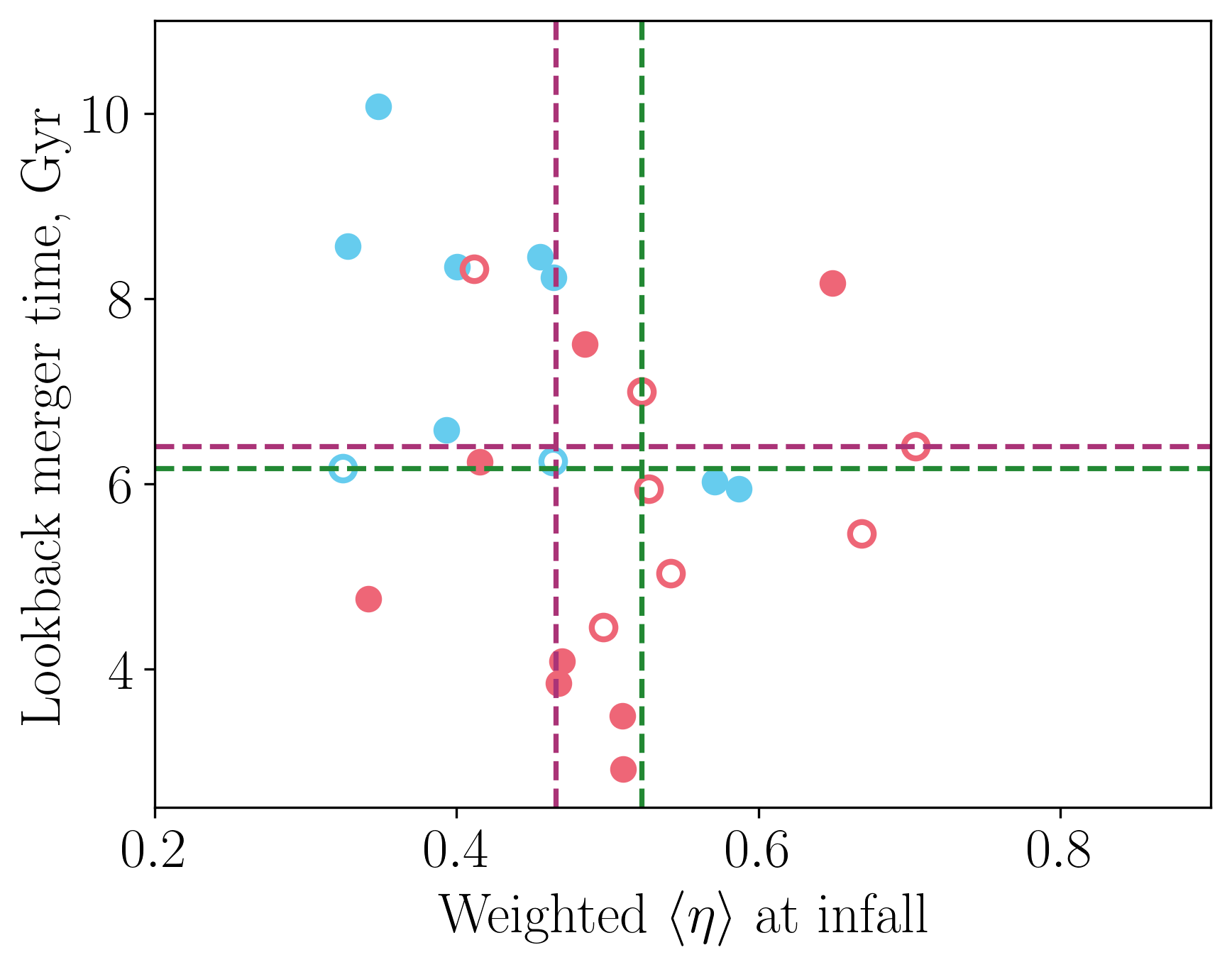}
&
\includegraphics[width=0.4108\textwidth]{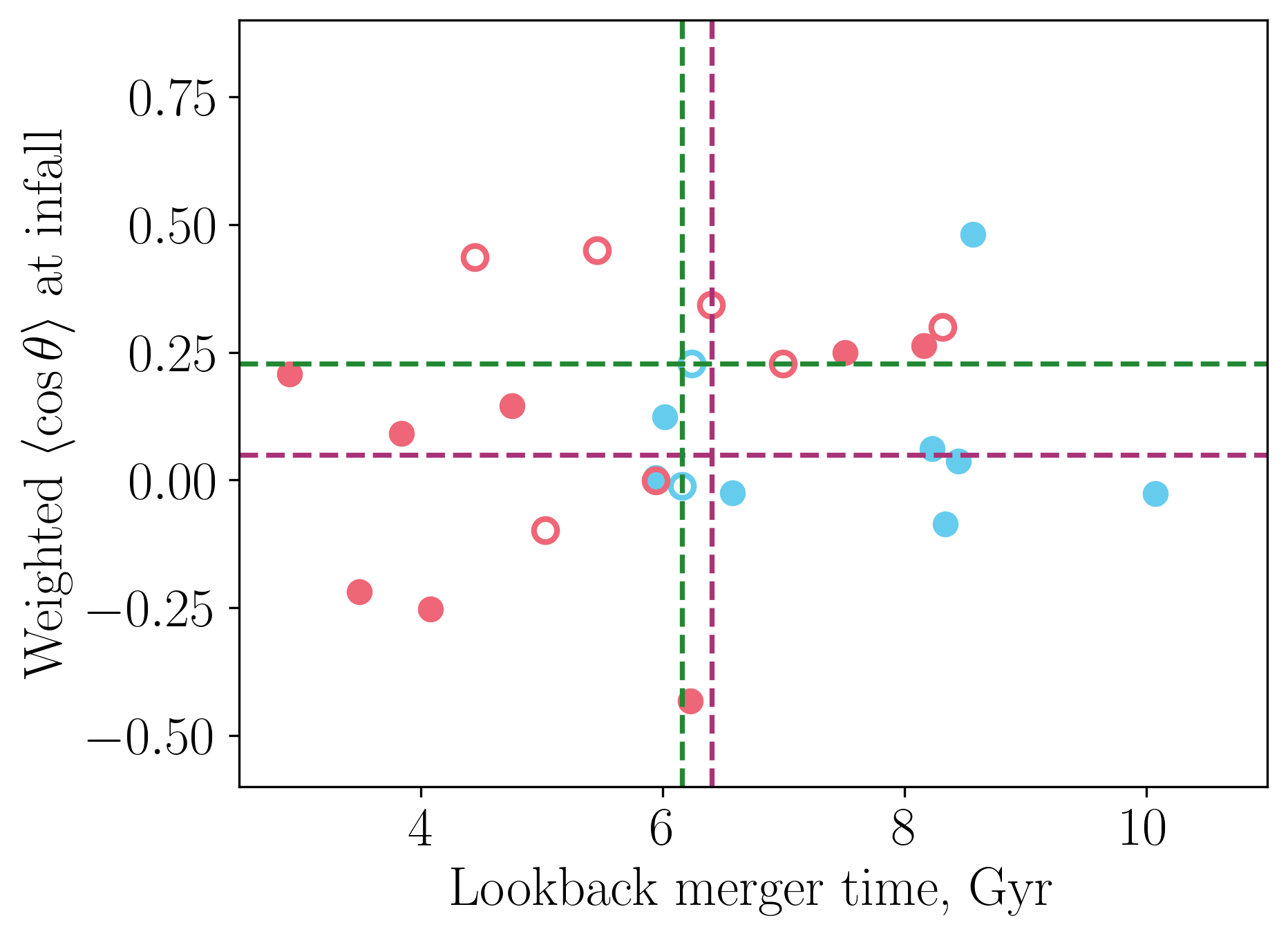} \\
    (c) & (d) 
    \end{tabular}
\caption[]{The upper row (a, b) shows scatter plots of weighted infall mass ratios of satellites (defined as in Equation~\ref{eq:M_ratio_avg}) against the orbital parameters $\eta$ and $\cos \theta$. The lower row (c, d) shows scatter plots of average lookback merger times (defined as described in Section~\ref{sec:res:z0}) against the orbital parameters. Note that the median values of mass ratios and merger times for haloes with and without disc flips are nearly identical. However, haloes with GES are typically formed earlier. [The legend is the same as in Figure~\ref{fig:eta_angle}].}
\label{fig:eta_angle_Mratio_mt}
 \end{figure*}

Having established that the orbital properties of satellites are important for the emergence of GES-like substructures and the occurrence disc flip events, we now ask whether mass ratios and merger times -- which were found to have a lesser impact on $v_{\phi}$ in Section~\ref{sec:res:z_evol} -- are also important for these events. To investigate this, we produce scatter plots in Figure~\ref{fig:eta_angle_Mratio_mt} similar to those in Figure~\ref{fig:eta_angle}, but showing infall mass ratios at infall (top row) and average lookback merger time (bottom row) against circularities (left panels) and misalignment angle (right panels). Notably, the medians of mass ratios for haloes with and without disc flips are essentially identical, indicating that the mass ratio of accreting satellites does not determine whether or not a disc flip occurs. The medians of merger times for haloes with and without disc flips are not identical, but remain very close to each other. This implies that the orbital parameters $\cos \theta$ and $\eta$ are more important in determining the disc flip event.

A further notable result from panels (c, d) of Figure~\ref{fig:eta_angle_Mratio_mt} is that haloes with GES-like substructures have distinctly earlier average merger times. While Auriga haloes were labelled as having a GES-like substructure by \citet{Fattahi_GES} purely based on the analysis of stellar haloes at $z_{\mathrm{red}}=0$, this is a reassuring result, as the GES merger event in the MW is currently thought to have occurred approximately $8-11$ Gyr ago \citep{Belokurov_GES, Helmi_GES}. Furthermore, the earlier formation of haloes with GES-like substructures likely contributes to the trend of slower $v_{\phi}$ in earlier-forming haloes seen in Figure~\ref{fig:vphi_radial_mergeT} and panel (a) of Figure~\ref{fig:acc_mt_Mratio_Zevol}, since 5 out of 7 haloes with an average merger time of $z_{\mathrm{red}}>1$ host a GES-like substructure.

\section{Discussion}
\label{sec:disc}

Following the findings presented in Section~\ref{sec:res}, we discuss our results and their implications in the context of previous studies on satellite anisotropy and halo rotation.

\subsection{Satellite anisotropy}
\label{sec:disc:anis}

The anisotropy of satellite distributions around MW-mass host haloes in numerical simulations has been the focus of several studies \citep[e.g.][]{Libeskind05, Zentner05, Deason11, Kang15, Shao21}.  Most studies find that the brightest satellites (i.e. those with the largest baryonic or stellar masses) are preferentially aligned with the host halo's major axis, reflecting the anisotropic accretion of matter along cosmic filaments. The alignment between satellite galaxies and the filament axis was also confirmed in the SDSS catalogue by \citet{Tempel15} and \citet{Peng20}. By extension, the debris from disrupted satellites (i.e. the stellar halo) may also exhibit a preferred orbital direction. This direction is generally expected to be prograde with respect to the spin of the host dark matter halo and stellar disc, as all components are fed by the same large-scale filamentary structure. However, there remains considerable scatter in the alignment between dark matter haloes, stellar discs, and satellite orbits, likely driven by complex and stochastic accretion histories.

The anisotropic satellite accretion is consistent with our results. The weighted average cosine of misalignment angle $\langle \cos \theta \rangle$ for all Auriga haloes in our sample (black dashed curve in panel (a) of Figure~\ref{fig:eta_theta_kde}) exhibits a non-uniform distribution with a peak around $\approx 0.2$ (roughly corresponding to $\theta \approx 80 \degr$), reflecting the tendency of satellites to be more aligned (i.e. more prograde) with the stellar disc than anti-aligned (i.e. more retrograde), albeit with considerable scatter. Thus, it is unsurprising that the disrupted satellites (i.e. the stellar halo) exhibit a weak prograde rotational signal.

\subsection{Dark matter and stellar haloes}
\label{sec:disc:halo}
In Figure~\ref{fig:halo_dm_corr}, we demonstrate a strong correlation between the rotation of stellar and dark matter haloes, suggesting a possibility of a common origin. Although the rotational velocity $v_{\phi}$ of dark matter haloes is not widely discussed in the literature, the temporal evolution of the dimensionless DM halo spin parameter $\lambda$ \citep{Peebles69} has been studied extensively. We emphasise that $\lambda$ and $v_{\phi}$ are not equivalent: $\lambda$ is a rescaled measure of angular momentum rather than a velocity. Nevertheless, studies of the spin parameter provide useful context for interpreting our results. Previous works \citep[e.g.][]{Bullock01, Vitvitska_DM_AM, Hetznecker06} have shown that $\lambda$ is sensitive to the merger history of the host halo, with mergers effectively regulating its angular momentum. Mergers arriving from similar directions can build up a coherent prograde spin, whereas more isotropic accretion tends to cancel angular momentum contributions, leading to a slower net spin. This supports our interpretation that $v_{\phi}$ is shaped by the host’s accretion history, particularly the anisotropy of satellite infall. The connection between the rotation of stellar and dark matter haloes is also consistent with numerical simulations by \citet{Obreja22}, who find correlated angular momentum magnitudes for the two components.

\subsection{GES}
\label{sec:disc:GES}
In Section \ref{sec:res:GES_flips} we showed that both stellar and DM haloes exhibit slower rotational velocity $v_{\phi}$ if they host a GES-like substructure at the present day. This is consistent with the findings of \citet{Dillamore_halo_spin}, who showed that DM haloes exhibiting GES-like features have a median dimensionless spin parameter lower by a factor of 1.7 compared to systems without a GES. Figure~\ref{fig:eta_theta_kde} shows that haloes with GES-like substructures have satellites most tightly concentrated around $\langle \cos \theta \rangle=0$ (i.e. perpendicular orientation) and the lowest $\langle \eta \rangle$ (i.e. most radial orbits) among other halo categories. The dominant progenitor in these systems therefore arrives on a near head-on trajectory, whereas in haloes without GES, satellites are on average --- albeit with significant scatter --- more aligned with the disc.
 
Satellites arriving on orbits more aligned with the disc contribute more coherently to the build up of the net rotation of stellar halo. This explains the difference in $v_{\phi}$ between haloes with and without GES. The slower rotation of haloes with GES is also consistent with the characterisation of the GES in the MW as a highly radially anisotropic structure \citep[][]{Belokurov_GES, Helmi_GES}. Furthermore, haloes with GES-like substructures typically exhibit a more quiescent merger history (as reflected in the average lookback merger time of stellar haloes in panels (c,d) of Figure~\ref{fig:eta_angle_Mratio_mt}). This implies that the relatively slow rotation associated with a dominant progenitor leading to a GES-like substructure is unlikely to be altered by later mergers, even if these are more aligned with the stellar disc.

\subsection{Disc flips}
\label{sec:disc:flips}
Finally, we discuss the importance of disc flips for the rotation of stellar and dark matter haloes. To avoid the ambiguity, we remind the readers of our definition of disc flips: a change in the angular momentum axis orientation by more than $90 \degr$, measured over all in-situ stellar particles within $r<30$ kpc. As shown in Figures~\ref{fig:eta_angle} and~\ref{fig:eta_angle_Mratio_mt}, systems that have undergone disc flips also typically exhibit lower $\langle \cos \theta \rangle$ (i.e. progenitors more misaligned with the disc) and lower $\langle \eta \rangle$ (i.e. more radial orbits), which naturally results in a weaker prograde rotational signal with respect to the disc (as seen in panel (b) of Fig.~\ref{fig:acc_dm_GES_flip_Zevol}) -- similar to what is seen for haloes with GES-like substructures. Since 8 out of 10 haloes with GES also experienced a disc flip it is tempting to simply attribute this to the influence of the dominant progenitor that led to emergence of a GES-like substructure. However, 8 additional haloes had a disc flip without GES, and all haloes with disc flips exhibit slower $v_{\phi}$ than haloes without disc flips in panel (b) of Figure~\ref{fig:acc_dm_GES_flip_Zevol}. Thus, the influence of disc flips on $v_{\phi}$ cannot be attributed solely to one dominant merger.

To place disc flips in a broader context, we briefly address previous works on this phenomenon. \citet{Dodge_disc_tilt} found that stellar disc flips are common, with more massive mergers producing more rapid reorientations of the angular momentum vector, although minor mergers can contribute as well. \citet{Welker14} found that galaxies with a larger number of mergers tend to have their discs tilted more than those with fewer mergers. \citet{Dillamore22} found that 4 out 15 galaxies with GES-like features in cosmological simulations underwent a rapid disc flip following the merger with the dominant progenitor, with no dependence on merger mass ratio. Finally, \citet{Bell26} recently used the TNG-50 simulations to show that disc tilting is frequently driven by significant mergers, particularly the dominant (most massive) progenitor. They argue that this merger-driven reorientation leads to a present-day alignment between the galaxy disc and the stellar halo, which is often dominated by the debris from the most massive progenitor. However, they caution that the Milky Way may be atypical in this respect, likely due to the early, radial nature of its dominant GES merger. In this work, we consider 25 Auriga haloes, and 10 of these host GES-like substructures. Thus, we probe a more limited (but perhaps more MW-specific) range of accretion histories compared to the TNG-50 sample.

Direct comparison of our results and those in the literature is not straightforward, as different definitions of disc flips or tilts were employed. Our results suggest that disc flips are more common in haloes whose accreted satellites follow more misaligned and more radial orbits, independently of the progenitor mass or merger time. We suggest that this can arise via two possible scenarios, each echoing a case discussed above: a single dominant misaligned progenitor (e.g. haloes with GES-like substructures and disc flips, analogous to the scenario described by \citealt{Dillamore22}), or the continued accretion of multiple satellites along a direction that is consistently misaligned with the disc (e.g. haloes without GES but still exhibiting disc flips, analogous to the scenario described by \citealt{Welker14}). In both cases, the resulting stellar halo has weaker net rotation. When the stellar disc undergoes a flip, its angular momentum axis evolves rapidly over time, while previously accreted halo stars, which have much longer dynamical times, are slower to adjust. This leads to a persistent misalignment between the stellar halo and the disc, reducing the coherence of rotation in the disc frame. As a result, systems that experience disc flips exhibit systematically weaker stellar halo rotation than those with a more stable disc orientation.

\section{Conclusions}
\label{sec:concl}
The MW stellar halo has been reported to have slow net rotation based on \textit{Gaia}+SDSS data \citep{Deason17} and, more recently, with \textit{Gaia}+DESI data \citep{Li26}. The origin of this rotation and the reason for its small magnitude ($|v_{\phi}|<25$ km s$^{-1}$) remain unexplained. We have analysed analogues of the MW in the Auriga simulation suite to investigate what factors lead to faster or slower stellar halo rotation. Our findings are summarised as follows.
\begin{enumerate}
    \item The Auriga haloes have significant disc-like structures of accreted stars, first discovered by \citet{Gomez_exs_discs} and termed `ex-situ discs'. We find that (at least partial) exclusion of such structures by applying spatial height cuts to accreted stars decreases the net $v_{\phi}$ in Auriga stellar haloes.
    \item Stellar haloes exhibit prograde net rotation with respect to stellar discs. We attribute this to the anisotropic accretion of satellites arriving from a preferential direction typically aligned with the host major axis.
    \item Haloes that host a GES-like substructure at $z_{\mathrm{red}}=0$ exhibit slower $v_{\phi}$ than haloes without such a substructure. This is likely because the dominant progenitor in systems with GES typically arrive from a head-on direction on a more radial orbit.
    \item Earlier-forming haloes (i.e. those with an average merger time of satellites at $z_{\mathrm{red}}>1$) exhibit slower rotation than later-forming haloes. This is likely a consequence of the previous point:  5 out of 7 earlier-forming haloes also host a GES-like substructure.
    \item Haloes that underwent a disc flip event (defined as a change in the stellar disc angular momentum vector orientation of $\geq 90 \degr$) exhibit slower rotation than haloes without disc flip events. We suggest that this is due to a persistent misalignment between stellar discs and stellar haloes in such systems: stellar haloes have much longer dynamical times and cannot adjust as quickly to the reoriented discs.
    \item The weighted average ratio of the satellite bound mass to the host bound mass at infall was not found to have a significant impact neither on the stellar halo net rotational velocity $v_{\phi}$, nor on the occurrence of disc flips. 
    \item A correlation in the net rotation of dark matter and stellar haloes is found at the present day. At redshifts up to $z_{\mathrm{red}}=2.5$ (the highest redshift considered in our analysis), the dark matter halo rotation is affected by the presence of GES-like substructures and disc flips in the same manner as the stellar halo rotation. These findings suggest a possible common origin for the rotation of both types of haloes. We reiterate that anisotropic accretion from a preferential direction is likely the main source of net rotation, as both the DM halo and satellites (which form the stellar halo upon their disruption) are fed by the same large-scale filaments.
\end{enumerate}

The slow net rotation of the MW stellar halo is consistent with our Auriga simulation results, given the presence of the GES substructure and a relatively quiescent merger history. Based on our findings, we hypothesise that the MW likely experienced a disc flip during its formation history, and that its dark matter halo, like its stellar halo, is very slowly rotating.

\section*{Acknowledgements}
We thank an anonymous referee for providing useful comments that
helped improve the paper. We have used simulations from the Auriga Project public data release \citep{Auriga24}. We thank Songting Li for providing observational data points of the MW stellar halo velocity field. We acknowledge support from the Science and Technology Facilities Council (STFC) [grant number ST/X001075/1]. KB is supported by an STFC PhD studenthip UKRI1738. FF is supported by a UKRI Future Leaders Fellowship (grant no. MR/X033740/1). TT is supported by an STFC PhD studentship 2876865. AF is supported by a Swedish Wallenberg Academy Fellowship. This work used the DiRAC@Durham facility managed by the Institute for Computational Cosmology on behalf of the STFC DiRAC HPC Facility (www.dirac.ac.uk). The equipment was funded by BEIS capital funding via STFC capital grants ST/P002293/1, ST/R002371/1 and ST/S002502/1, Durham University and STFC operations grant ST/R000832/1. DiRAC is part of the National e-Infrastructure. For the purpose of open access, the authors have applied a Creative Commons Attribution (CC BY) licence to any Author Accepted Manuscript version arising from this submission.
%The Acknowledgements section is not numbered. Here you can thank helpful colleagues, acknowledge funding agencies, telescopes and facilities used etc. Try to keep it short.

%%%%%%%%%%%%%%%%%%%%%%%%%%%%%%%%%%%%%%%%%%%%%%%%%%
\section*{Data Availability}
Detailed instructions on how to access the Auriga public data are available at \url{https://wwwmpa.mpa-garching.mpg.de/auriga/data.html}.
 
%The inclusion of a Data Availability Statement is a requirement for articles published in MNRAS. Data Availability Statements provide a standardised format for readers to understand the availability of data underlying the research results described in the article. The statement may refer to original data generated in the course of the study or to third-party data analysed in the article. The statement should describe and provide means of access, where possible, by linking to the data or providing the required accession numbers for the relevant databases or DOIs.



\bibliographystyle{mnras}
\bibliography{biblist} 

%%%%%%%%%%%%%%%%%%%%%%%%%%%%%%%%%%%%%%%%%%%%%%%%%%

%%%%%%%%%%%%%%%%% APPENDICES %%%%%%%%%%%%%%%%%%%%%

%\appendix



%%%%%%%%%%%%%%%%%%%%%%%%%%%%%%%%%%%%%%%%%%%%%%%%%%

% Don't change these lines
\bsp	% typesetting comment
\label{lastpage}
\end{document}